\documentclass[12pt]{article}
\usepackage{fullpage}
\usepackage{amssymb, amsthm, amsmath}
\usepackage{graphicx}
\usepackage[authoryear]{natbib}
\usepackage{bm}
\usepackage{subfig}
\usepackage{verbatim}
\usepackage{lineno}
\usepackage{times}
\usepackage{tabulary}
\usepackage{url}
\usepackage{geometry}
\usepackage{color}

\newcommand{\btheta}{ \mbox{\boldmath $\theta$}}
\newcommand{\bTheta}{ \mbox{\boldmath $\Theta$}}

\newcommand{\bbeta}{ \mbox{\boldmath $\beta$}}

\newcommand{\brho}{ \mbox{\boldmath $\rho$}}

\newcommand{\bxi}{ \mbox{\boldmath $\xi$}}
\newcommand{\bepsilon}{ \mbox{\boldmath $\epsilon$}}

\newcommand{\bd}{ \mbox{\bf d}}

\newcommand{\bx}{ \mbox{\bf x}}
\newcommand{\bX}{ \mbox{\bf X}}

\newcommand{\bZ}{ \mbox{\bf Z}}

\newcommand{\bY}{ \mbox{\bf Y}}
\newcommand{\bJ}{ \mbox{\bf J}}

\newcommand{\bs}{ \mbox{\bf s}}

\newcommand{\bu}{ \mbox{\bf u}}

\newcommand{\arginf}{{\mathop{\rm arg\, inf}}}

\newcommand{\beq}{ \begin{equation}}
\newcommand{\eeq}{ \end{equation}}
\newcommand{\beqn}{ \begin{eqnarray}}
\newcommand{\eeqn}{ \end{eqnarray}}

\begin{document}

	\begin{center}
		{\Large Computer model calibration based on image warping metrics: an application for sea ice deformation}\\\vspace{6pt}
		{\large \footnote{The first two authors contributed equally to this work} Yawen Guan\footnote{North Carolina State University}\footnote{The Statistical and Applied Mathematical Sciences Institute}, Christian Sampson\footnote{The University of North Carolina at Chapel Hill}\footnote{The Statistical and Applied Mathematical Sciences Institute}, J. Derek Tucker\footnote{Sandia National Laboratories}, Won Chang\footnote{University of Cincinnati},  Anirban Mondal,\footnote{Case Western Reserve University}, Murali Haran\footnote{Pennsylvania State University} and Deborah Sulsky\footnote{University of New Mexico}}\\
	\end{center}
	\begin{abstract}
			\noindent
			Arctic sea ice plays an important role in the global climate. Sea ice models governed by physical equations have been used to simulate the state of the ice including characteristics such as ice thickness, concentration, and motion. More recent models also attempt to capture features such as fractures or leads in the ice. These simulated features can be partially misaligned or misshapen when compared to observational data, whether due to numerical approximation or incomplete physics. In order to make realistic forecasts and improve understanding of the underlying processes, it is necessary to calibrate the numerical model to field data. Traditional calibration methods based on generalized least-square metrics are flawed for linear features such as sea ice cracks. We develop a statistical emulation and calibration framework that accounts for feature misalignment and misshapenness, which involves optimally aligning model output with observed features using cutting edge image registration techniques. This work can also have application to other physical models which produce coherent structures.	\vspace{12pt}\\

			{\bf Key words:} Arctic Sea Ice; Calibration; Emulation; Gaussian Process; Image Registration.
	\end{abstract}
	\section{Introduction}\label{s:intro}
{\color{black} Sea ice is frozen sea water that insulates the warmer ocean from the colder atmosphere in winter and provides a partial barrier to heat, moisture, and momentum transfer between the ocean and atmosphere, making it an important component in the Earth's energy balance. The ice cover waxes and wanes seasonally and at its maximum extent covers around 7\% of the Earth's surface and close to 12\% of the Global Ocean \cite{Weeks2010}. Cracks in the ice cover, called leads, occupy 1-2\% of the ice cover in winter but account for half of the heat flux from the ocean to the atmosphere \cite{Badgley1961}.}\\

A constitutive model, the elastic-decohesive model \citep{SchreyerEtAl2006} has been developed to describe crack formation in sea ice. The important aspect of this model is that the existence as well as the orientation and the opening or closing of a crack is directly represented. Parameters defining the material properties for the elastic-decohesive model were often chosen based on experiments \citep{Schulson2004} and some {\em in situ} observations \citep{CoonEtAl1998}. There has never been a more detailed study of the appropriateness of these parameters and whether other choices would better fit observations. \\

With new high resolution satellite data now available, unprecedented opportunities to improve our understanding of the interaction between dynamics and thermodynamics of sea ice, and to appraise new models are now possible. With this imagery, we are able to resolve long linear features in the pack ice that evolve with time that can generally be associated with fractures or leads. These leads are derived from the small-scale, time-varying deformation of the ice cover observed using the RADARSAT Geophysical Processing System (RGPS) \citep{Kwok2001}.\\

Thus, a fresh assessment, through calibration of the model parameters, is appropriate. {\color{black} The focus of this paper is to develop metrics suitable for comparing simulation and observations of these linear features in sea ice, and to use these metrics to calibrate the elastic decohesive model.} Calibration is the traditional statistical method to infer unknown input parameters for a model of a physical process from the corresponding observed output. The Bayesian framework \citep{Kennedy2001, Higdon2004CombiningFD} is a natural choice for calibration where the inference is based on the posterior distribution of the model parameters given the observed outputs. Here, the likelihood term in the posterior will include the non-linear physical forward model, via, the elastic-decohesive model, which is computationally expensive to solve numerically. Moreover, the likelihood needs to be evaluated at each iteration of the Markov Chain Monte Carlo (MCMC) sampling scheme, thus making the Bayesian calibration method computationally very challenging. Alternatively, a statistical emulator can be constructed using the simulation runs of the elastic-decohesive model at some suitably chosen design points. The emulator can then act as a surrogate, providing fast predictions of the simulation output at unsampled input values, with corresponding measures of uncertainty \citep{oakley2002, conti2009}. In practice, a Bayesian modularization approach \citep{Bay2007, Liu2009} is often used, in which a Gaussian process (GP) based emulator is fit only once, then the fitted emulator replaces the physical model inside the likelihood for Bayesian calibration. We adapt this in our proposed method. \\

There are two major challenges in applying the traditional emulation-calibration framework to linear features such as sea ice cracks. (1) In the traditional framework, the output from the elastic-decohesive model would be modeled as a realization from a GP; however, emulating the response surface with linear features using a GP is inadequate. Although the GP model is attractive for emulation because of its flexibility to fit a large class of response surfaces, it is not suitable for emulating a surface with sharp gradients and discontinuities. For model outputs with linear features, a simple GP emulator would tend to smooth out those features. (2) Generalized least-square metrics are implicitly used in the likelihood functions when fitting a GP emulator and carrying out calibration; however, such metrics are also flawed when used in the presence of sharp gradients and discontinuities. Because the linear features occupy only a small, usually lower dimensional, region of the domain, models that completely fail to predict these important features may actually score better by these metrics than those that simulate the dynamically important features albeit at a somewhat different location or orientation than observations indicate \citep{MassEtAl2002}. We propose a novel emulation-calibration framework, which utilizes cutting-edge image registration technique, to circumvent the above challenges in dealing with linear sea ice features. It involves optimally aligning high-dimensional model outputs to the observed features, then performing emulation and calibration based on the metrics provided from image registration. \\

Image registration, finding a correspondence between pixels across images,  is a standard problem in image processing (see \citep{goshtasby:2012} for a review).
Recently, the concept of \textit{warping} images for registration using metrics has become increasingly popular given the properties they provide (\citet{beg:2005}, \citet{tagare:2009}, \citet{vercauteren:2009}, \citet{xie:16}).
Image registration is usually posed as a variational problem where the objective function between two images $f_1$ and $f_2$ is of the form 
$$
g(f_1,f_2\circ\gamma) = \int_D |f_1(s) - f_2(\gamma(s))|^2\,ds + \lambda R(\gamma), ~~\gamma\in\Gamma
$$
where $|\cdot|$ is the Euclidean norm, $\gamma$ is the deformation, $R$ is a regularization to ensure smooth deformations, and $\Gamma$ is the space of relevant deformations. Recently, this metric has been applied to model emulation and calibration \citep{kleiber:2014}. 
However, the `metric' used in Kleiber's approach is not a proper metric in the image space under warping and can fail to properly calibrate when observation and simulation images are not properly aligned.\\

The underlying problem with the above formulation is that it does not provide a proper metric that is \textit{inverse consistent}. Here we propose a new method using the concepts of phase and amplitude of an image. 
The power of this method is that we have two metrics, first proposed by \citet{xie:16}, which are proper distances in the space of images. One, the phase energy, gives a measure of how much warping is required to register geometric features. The other, the amplitude energy, gives a measure of the pixel intensity differences, if any, after an optimal warping has been applied. The combination of the measures allows one to determine how well a model is reproducing geometric features or coherent structures. A low amplitude energy, for instance, implies there is a smooth deformation from one image to the other, meaning a geometric feature is likely captured but perhaps misaligned. The magnitude of the phase energy gives a measure of how misaligned the feature is. In contrast, a high amplitude energy implies the geometric features of the two images are inconsistent and are not captured by the model. Using these measures in our calibration scheme allows the emulator to better distinguish how different a model prediction {\it truly} is from an observation.\\

We propose to use these two informative metrics to train our emulator for use in calibration. This allows us to use fewer parameter settings for expensive model runs while providing a unique measure of model performance. The advantage of our approach is two-fold: (1) the alignment avoids penalizing predicted features that are misaligned and misshapen in calibration, and (2) emulation-calibration for scalar metrics is straightforward so that standard software can be applied directly. The rest of this paper is organized as follows, Section \ref{s:data} introduces the RGPS deformation data and the sea ice model that describes the physics of sea ice deformation. Section \ref{s:method} introduces image warping and the emulation-calibration framework based on the metrics provided by the warp. We illustrate the usefulness of our approach via a toy example and a perfect model experiment based on the sea ice model in Section \ref{s:simu}, and we apply the proposed method to the RGPS data and present the results in Section \ref{s:appli}. Some final remarks on the proposed approach are included in Section \ref{s:discussion}.

\section{Data}\label{s:data}
The fundamental quantity produced by RGPS is trajectories of marked sea-ice points. At an initial time, a set of points forming a square grid is located in a synthetic aperture radar (SAR) image. In images resulting from subsequent satellite passes, the original points are found again using area-based and feature-based tracking. The time separation between repeat observations is variable and is based on available coverage. The time interval between successive images is called the timestep. This procedure provides the trajectory of each point as these points move with the ice cover. Since the same set of points is tracked over a ice season, RGPS provides a densely sampled Lagrangian picture of the motion, similar to what would be obtained by drifting buoys. Secondary procedures in RGPS derive estimates of ice deformation. If the set of points in the original configuration is viewed as vertices of grid cells then the motion of the points determines the deformation of those cells. With this interpretation, grid quantities can be approximated that help provide a picture of the rate and type of ice deformation.\\
	
The RGPS deformation product consists of cell area changes and spatial derivatives computed using the ice displacements at the cell vertices. Each cell has a unique identifier, the time of its creation, the map coordinates of the cell center, the displacement of the cell center between this and the last observation, the timestep and the derivatives of displacement increments. Since leads are formed by cracks in the ice, the displacement associated with a lead is discontinuous. The article by \cite{CoonEtAl2007} introduced the idea of separating the displacement into the sum of a continuous and discontinuous part. A kinematic analysis of RGPS deformations was performed assuming that the predominant term was the jump discontinuity in displacement. This procedure fit a jump in displacement, or a crack, in each RGPS cell that could best account for the observed deformation.  This fitting procedure provides the orientation and amount of normal and shear opening of the crack.
{\color{black} Details on this fitting processes can be found in \cite{porsec2011}. Figure~\ref{fig:RGPSCal}(b) shows an example of applying the procedure to a region of the Beaufort Sea to plot the magnitude of the predicted jump in displacement.\\
	
Observations and simulations of a region of the Beaufort Sea between Banks Island and Point Barrow during the period from Feb. 23 - Mar 11, 2004, are analyzed in this paper. During this period, daily observations are available on a 10 km square grid.}  Specifically, the domain of interest is described by the rectangle, $D=[-2345,-1505]\times [-260, 730]$ in SSM/I projection kilometers. Part of this region is land and the remainder is ice covered. The 16 days of RGPS observations are processed using the kinematic algorithm to obtain a jump in displacement, $[\mathbf{u}](\bs)=u_n \mathbf{n}+u_t \mathbf{t}$, $\bs = (s_1,s_2) \in D$, in each RGPS cell.  This jump in displacement represents a crack with a normal to the crack given by $\mathbf{n}=(\cos \alpha(\bs), \sin \alpha(\bs))$ and a tangent to the crack given by $\mathbf{t}=(-\sin \alpha(\bs), \cos \alpha(\bs))$ in the plane of the ice. Thus, the orientation of the crack is determined by $\alpha$, the angle that the normal makes to the $s_1$-axis.  The size of the crack is obtained from the normal and tangential components of the jump vector, $u_n$ and $u_t$, respectively. We threshold the jump in displacement and drop fractures smaller than 0.4 km.\\
	
Simulations using the material-point method (MPM) for this problem are described in \cite{SulskyPeterson2011} and output the same jump vector, $[\mathbf{u}]$, as that which can be derived from the RGPS data.  To summarize, the simulations are performed on the same 10 km grid as the observations, initially with four material points per cell. Land is modeled with rigid, immobile material points. Boundaries of the rectangular region intersecting ice are given a prescribed velocity in agreement with the RGPS observations.  At the start of the calculation the ice is at rest with a uniform average thickness. The thickness distribution producing the average thickness is assumed to be parabolic with a maximum at $h=3$~m and equaling zero at $h=0$~m and $h=6$~m. Six hour wind fields from the National Center for Environmental Protection and National Center for Atmospheric Research (NCEP/NCAR) reanalysis \citep{Kalnay_etal_1996} are used to determine the wind drag. Additionally, six hour air temperature, specific humidity, long wave flux, and short wave flux from the NCEP/NCAR reanalysis are used for the ice column temperature calculation. Details of the thermodynamic model are also in \cite{SulskyPeterson2011}. The ocean currents are updated daily using the output from an ocean model (MITgcm, \cite{Marshall_etal_1997}) run independently from the ice simulation. Nominal values of sea surface temperature and salinity of ${-1.8^\circ}$C and 32 ppt are used throughout the domain for the calculation.\\
	
The elastic-decohesive constitutive model used in the simulations is based on two elastic parameters, Young's modulus, $E$, and Poisson's ratio ($\nu=0.36$), and five independent decohesive parameters.  There are three strength parameters: tensile failure strength ($\tau_\text{nf}$), shear failure strength ($\tau_\text{sf}$), and compressive failure strength ($f_c'$); plus the shear magnification factor set to 4, and the opening parameter, $u_0$, set to 0.4~km. For the calibration, $\tau_\text{nf}$ is chosen uniformly in the range 10-50~kPa and the shear strength in the range 10-150 kPa. These choices allow the shear strength to be both larger and smaller than the tensile strength.The compressive strength is set to $5\tau_\text{nf}$ and the Young's modulus to $E = 10f_c'$. Keeping these ratios means that $u_0$ would not need to be changed to maintain stability of the algorithm and avoid snapback. Note that these strength ranges are smaller than what has been seen in experiments (eg. \cite{TimcoWeeks2010}), but are larger than what is typical for large scale simulations. The fact that sea ice on a large scale might be effectively weaker could be attributed to features such thermal bending fractures that appear on a scale of about 1~km, or other impurities and weaknesses that would not show up on a laboratory scale, but are smaller than typical simulation scales.\\

Fractures are initialized in the simulations using the kinematic analysis of the RGPS data over one day (23 February). Depending on deformation in a cell the calculated jump in displacement can reflect an opening mode ($u_n>0$) or a closing mode ($u_n<0$). An opening mode is initialized as a crack with the amount of opening prescribed by the kinematic analysis. The discontinuities in a closing mode are initialized as cracks with the calculated direction and tangential jump in displacement, but with zero normal jump in displacement. After the initialization, the simulation is run for 15 more days. The initial fractures may open or close, and other fractures may be created and evolve.  The resulting predicted values of $u_n$, $u_t$ and $\alpha$ are compared to the kinematic analysis of RGPS data.

\section{Statistical Method}\label{s:method}

We propose a novel emulation and calibration framework based on proper distance metrics defined on the image space, this allows us to properly account for the geometric differences between model outputs and observations. In this section, we first introduce the concept of phase and amplitude distances provided by image warping, where the first one quantifies the amount of geometric deformation required to correct for misaligned and misshapen model errors, while the latter quantifies the remaining model error after optimal alignment. We then describe how to construct an emulator and carry out calibration based on these two scalar metrics. This simplifies the high-dimensional image model outputs to scalar responses and significantly reduces the complexity of the statistical emulator and parameter inference procedure. 
	
\subsection{Image Warping}\label{s:warp}
Our approach to image warping is to use the concept of phase and amplitude of an image that was introduced by \citep{xie:16}. 
This work has been motivated by the the same concept developed in the work on shape analysis of objects and phase-amplitude modes in functional data (see \citet{srivastava-fda}, \citet{marron2015}, and \citet{tucker-wu-srivastava:2013} and references therein for a review of the concept).
A deformation is defined to be a change of only the phase of an image (where the pixels are located) and not the amplitude. 
This view point then sets up the registration problem. \\

Let the set of all images be $\mathcal{F} = \{f: D \rightarrow \mathbb{R}^n | f \in C^\infty(D)\}$ of the images in the domain $D$ that map to pixel values in $\mathbb{R}^n$.  {\color{black} Here we require $n\geq \dim(D)$ for the theory presented in this section. An example of a suitable image is that of an RGB image. In this case, $D=\mathbb{R}^2$ and $n=3$,  the space of red, green, and blue intensities at a given $s_1,s_2\in D$. In the case of a gray scale image, where $f(s_1,s_2):D\subset \mathbb{R}^2 \rightarrow \mathbb{R}$, one may instead consider an analogous image using the gradient of $f$ such that $F(s_1,s_2)=(\partial_{s_1} f,\partial_{s_2} f)$ or a combination of intensity and gradient such as $F(s_1,s_2)=(f,\partial_{s_1} f, \partial_{s_2} f)$. We discuss our particular choices for image type in Section \ref{s:experiment_ice}.}
Also let $\Gamma = \textnormal{Diff}^+(D)$ be the set of of all orientation-preserving diffeomorphisms that preserve the boundary of $D$.
It can be shown that $\Gamma$ is a group that contains an identity and inverse element.
The registration of two images $f_1$ and $f_2$ is expressed as finding $\gamma^*$ such that
\begin{equation}
	\gamma^* = \arginf_{\gamma\in\Gamma} ||f_1 - f_2 \circ \gamma||
	\label{eq:registration_prob}
\end{equation}
where $||f||$ is the $\mathbb{L}^2$-norm, $||f|| = \sqrt{\int_D |f(s)|^2\,ds}$.
{\color{black} However, (\ref{eq:registration_prob}) is not a proper Riemannian metric; specifically, it lacks the symmetry property.}


To overcome this problem \citet{xie:16} introduced the q-map, where for an $f\in\mathcal{F}$, we have a mapping $Q:\mathcal{F}\rightarrow L^2$ for any $s\in D$,
\begin{equation}
	Q(f)(s) = \sqrt{a(s)}f(s)
\end{equation}
where $a(s) = ||\bJ f(s)||_V$  and $\bJ f(s)$ is the Jacobian matrix of $f$ at $s$. 
The term $a(s)$ is known as the generalized area multiplication factor of $f$ at $s$. The use of the q-map provides a distance that is a proper metric on the space of images for the registration problem
\begin{equation}
	\gamma^* = \arginf_{\gamma\in\Gamma} ||q_1 - (q_2, \gamma)||
	\label{eq:registration_prob2}
\end{equation}
where $q = Q(f)$ and $(q,\gamma) = \sqrt{\textnormal{det}\bJ~ \gamma}(q\circ\gamma)$.
For proofs of the metric properties of (\ref{eq:registration_prob2}) the reader is referred to \cite{xie:16}.
This registration problem can then be solved using gradient descent.\\

The gradient descent is computed iteratively over $\Gamma$, where at each iteration the directional derivative is taken on the tangent space of $\Gamma$ at the identity element ($T_{\gamma_{id}}(\Gamma)$). 
The directional derivative is computed using an orthonormal basis expansion on $T_{\gamma_{id}}(\Gamma)$ and details are given in \cite{xie:16}. The current implementation of the gradient decent scheme is sensitive to step size and the number of basis functions used to represent $\gamma^*$ in the orthornomal expansion. These can be tuned, however, for the specific problem.\\

This registration problem then gives rise to the definition of amplitude distance between two images
\begin{equation}
	d_a(f_1, f_2) = \inf_{\gamma\in\Gamma}||q_1 - (q_2, \gamma)||.
\end{equation}
We can then also define a distance on the amount of deformation required to register the images. Since the space $\Gamma$ does not have a simple Riemannian geometry or one cannot be found through a transformation as in \citep{tucker-wu-srivastava:2013} we will use an extrinsic approach. We will use the q-map of the warping function and the resulting $\mathbb{L}^2$ metric. This gives us a nice metric for computing the distance, but further statistical analysis becomes hard, such as principal component analysis, due to the fact that the inverse mapping is not available in closed form.  We will call this the phase distance and it is defined as
\begin{equation}
	d_p(\gamma) = ||q_{\gamma{id}} - q_\gamma||
\end{equation}
where $q_\gamma = Q(\gamma)$. 
This distance measures how far the computed warping is from the identity warping. 
We will use these two distances in the calibration solution discussed later on.

\subsection{Statistical Emulator}\label{s:emu}
The sea ice model represents the formation and evolution of leads(cracks) resulting from the complex interactions among input processes. Due to the computational cost, we have a limited number of model runs. A standard approach is to replace the computer model by a fast surrogate statistical model often called an \textit{emulator}. Typically, the first step is to obtain model outputs at a set of design points, then construct a statistical model such as a Gaussian process to interpolate these model outputs \citep{oakley2002, Higdon2004CombiningFD, conti2009}.\\
	
Instead of interpolating model outputs, we use an emulator to interpolate the amplitude and phase distances, $d_a$ and $d_p$, provided by image warping. We let $\bY(\btheta) \in \mathcal{F}$ denote the model output corresponding to input parameter $\btheta \in \bTheta\subset\mathcal{R}^d$ for an integer $d\ge1$, then the distance metrics are computed by warping the model output to the reference image $ \bZ \in \mathcal{F}$ and $\bd (\bZ, \bY(\btheta)) = (d_a(\bZ,\bY(\btheta)), d_p(\gamma))^T$, where $\gamma = \arginf_{\gamma\in\Gamma} ||q_{Z} - (q_{Y}, \gamma)||$. For brevity, we write this as $\bd(\btheta)$, and implicitly assume in the remainder of the paper that it depends on the numerical model $\bY$ as well as the reference image $\bZ$. In the proposed emulation and calibration framework, we use the observation $\bZ$ as the reference image. 
Let $\btheta_1,\dots,\btheta_N \in \bTheta$ denote a set of design points and let $\bY(\btheta_i) \in \mathcal{F}$ denote the model outputs obtained at input parameter $\btheta_i$ for $i=1,\dots,N$. Each of these model outputs are warped to match the reference image and we denote them as $\bd=(\bd_1,\dots,\bd_N)^T$.\\

We construct a Gaussian process emulator $\boldsymbol{\eta}(\btheta, \bd)$ with mean $\mu=(\mu_a, \mu_p)^T$ and covariance function $C(\cdot,\cdot)$ for the distance metrics $\bd(\btheta)$. For the mean we use a linear function $\mu_k = \bx^T\bbeta_k$ for both $k = a \text{ and } p$, where $\bx$ are the covariates and defined as the input parameter $\btheta$ plus an intercept term, and $\bbeta_k$ is a $d+1$-dimensional vector of regression coefficients.  The covariance function models the dependence between every pair of metrics in terms of their distance in the parameter space. For example, we have used the exponential covariance function in our simulated example and data application. It is a commonly-used, separable, and stationary covariance function with the following form,
\beq
\begin{aligned}
\text{Cov}\left( d_k(\btheta),d_k(\btheta')\right) &= C(\btheta,\btheta'; \bxi_k) \\
&= \sigma^2_k\exp\left(-\sum_{j=1}^{d} \frac{|\theta_j-\theta_j'|}{\rho_{kj}}\right) + \tau^2_k1(\btheta=\btheta'),\text{ for } k = a,p,
\end{aligned}
\eeq
where $\rho_{k1},\dots,\rho_{kd}$ are the range parameters, $\sigma^2_k$ is the partial sill, and $\tau^2_k$ is the nugget. The range parameters control the rate of the correlation decay with distance for each dimension of the input space. Here we use $\bxi_k = (\brho_k, \sigma^2_k, \tau^2_k)$ to denote the parameters for the statistical emulator, to distinguish them from the physical model parameters $ \btheta $. We consider independent Gaussian processes for amplitude and phase distance since these two metrics are uncorrelated and hence we set $\text{Cov}\left( d_a(\btheta),d_p(\btheta')\right) = 0$. {\color{black} Other covariance funtions can be used here; we suggest performing cross-validation to select the appropriate covariance function.}  \\
	
The emulator is then constructed based on the metrics evaluated at the chosen design points using Latin hypercube sampling (LHS) \citep{mckay1}. Let $\bX$ be an $N\times (d+1)$ co-variate matrix and $\bd_k=(d_k(\btheta_1),\dots,d_k(\btheta_N))^T, k=a,p$ be the vectors of amplitude and phase distance metrics evaluated at the design points. The Gaussian process emulator described above specifies $\bd_a$ and $\bd_p$ as the following independent multivariate normal random variables:
\begin{equation*}
\bd_k \sim MN\left(\bX\bbeta_k,\Sigma(\bxi_k)\right)~\mbox{for k = a and p},
\end{equation*}
where $\Sigma(\bxi_k)$ is the $N\times N$ covariance matrix with elements $\left\lbrace\Sigma(\bxi_k)\right\rbrace_{ij}=C(\btheta_i,\btheta_j;\bxi_k)$. We estimate the emulator parameters $\bbeta_k$ and $\bxi_k$ by maximizing the likelihood given by the above model and denote the resulting maximum likelihood estimators (MLE's) as $\hat{\bbeta}_k$ and $\hat{\bxi}_k$. For any new input parameter setting $\btheta\in\bTheta$, the emulator predicts its corresponding distance metrics through the conditional normal distribution of $d_a(\btheta)\mid \bd_a$ and $d_p(\btheta)\mid \bd_p$, using the MLE's $\hat{\bbeta}_k$ and $\hat{\bxi}_k$ as plug-in estimates. This predictive process serves as the surrogate model in the following section on model calibration.

\subsection{Model Calibration}\label{s:cali}
We use $\bZ_d=(d_a,d_p)^T = (0,0)^T$ as the observation for calibration, since we use the observed image $\bZ$ as the reference, and registering an image to itself results in zero amplitude distance and zero phase distance. We assume that the data model for $\bZ_d$ is  
\beq
\bZ_d = \eta(\btheta^\ast,\bd) + \bepsilon,
\eeq
where $\btheta^\ast$ is the `best' parameter setting that yields $\bd(\btheta^\ast)=\mathbf{0}$. The data-model discrepancy  $\bepsilon = (\epsilon_a,\epsilon_p)^T$ is assumed to follow a truncated normal distribution with the following density:
\beq
\epsilon_k=
\begin{cases}
\frac{\phi(\epsilon_k / \psi_k)}{1-\Phi(\epsilon_k/\psi_k)},& \mbox{if }\epsilon_k\ge0,\\
0 & \mbox{if }\epsilon_k<0,
\end{cases}
\eeq
with the discrepancy standard deviation $\psi_k>0$ for $k=a,p$ and the probability and cumulative density functions $\phi$ and $\Phi$ for the standard normal distribution. We have chosen a truncated normal distribution with lower bound zero since we assume that the distance metrics for model outputs cannot take lower values than those for the  observational data $\bZ$.\\

We choose to adapt the standard Bayesian calibration framework for parameter inference as it provides more detailed information on input parameter uncertainty, in the form of posterior densities \citep{Kennedy2001}. Each input parameter in $\btheta$ receives a uniform prior with lower and upper bounds determined by its physical limits. Each discrepancy variance parameter $\psi^2_k$ receives an informative inverse gamma prior with a shape parameter $a_k = 20$ and a scale parameter $b_k = a_k d_{k,10^{th}}^2$ for $k=a,p$, where $d_{k,10^{th}}$ denotes the $10^{th}$ quantile of the distance metrics $\bd_k$ obtained at the design points. This prior encourages the resulting full conditional posterior for the standard deviation $\psi_k$ to have its mode around $ d_{k,10th}$, so that roughly 10 percent of the existing model runs at the design points are covered by the 95\% credible interval. {\color{black} Informative priors are chosen for the variance parameter because we have only one observation for calibration; this inverse problem is non-identifiable without strong priors}. The posterior samples for $\btheta$, $\psi^2_{a}$ and $\psi^2_{p}$ are obtained from Markov chain Monte Carlo using the Metropolis-Hastings algorithm.

\section{Experiments}\label{s:simu}
We conduct two experiments to illustrate the usefulness of the proposed method. In the first experiment, we generate model outputs from applying a diffeomorphism on a template that represents a feature similar to a crack. The second experiment is based on the sea ice model outputs, where one run of the model is randomly chosen to be the synthetic observation. The experiments establish proof of concept, the first in a simplified setting and the second with a complex geometry that is consistent with the real data application. {\color{black} The experiments also establish the applicability of this method to discrete data sets, even though the theory itself is developed in a space of smooth functions.} In both experiments, we compare calibration results using the Euclidean distance, phase distance, amplitude distance, and both phase and amplitude distance combined. 
	
\subsection{Diffeomorphism Warp}

We illustrate the calibration method using a toy example with geometric features similar to the ice opening features resulting from ice deformation. The mathematical model used here applies a boundary preserving differmorphism to a 2-dimensional surface {\color{black} on the unit square}, with input parameters $\btheta = (\theta_1,\theta_2)$ to define the map. {\color{black} This allows us to evaluate our calibration method on a data set where simulation differs from observation by an actual diffeomorphism, an ideal case.} Let $\bs=(s_1,s_2)$ denote the coordinate of an image pixel. The diffeomorphism $W: \mathbb{R}^2 \to \mathbb{R}^2 $ maps a 2-dimensional coordinate to a new coordinate that is also 2-dimensional. Let $ \bs'=(s_1',s_2')$ denote the new location after warp, then we have $\bs' = W(\bs) $ where $\bs'$ is given by,
\beq\label{eqn:warp}
\begin{aligned}
	s_1' &= s_1-2\theta_1s_2\sin(s_1)\sin(s_2)(\cos(\pi s_1)+1)(\cos(\pi s_2)+1)\\
	s_2' &= s_2+2\theta_2s_1\sin(s_1)\sin(s_2)\cos(\pi/2 s_1)\cos(3/2\pi s_2).
\end{aligned}
\eeq\\

\begin{figure}[hpt]
	\centering

	\subfloat[Template image]{\includegraphics[width=0.3\textwidth]{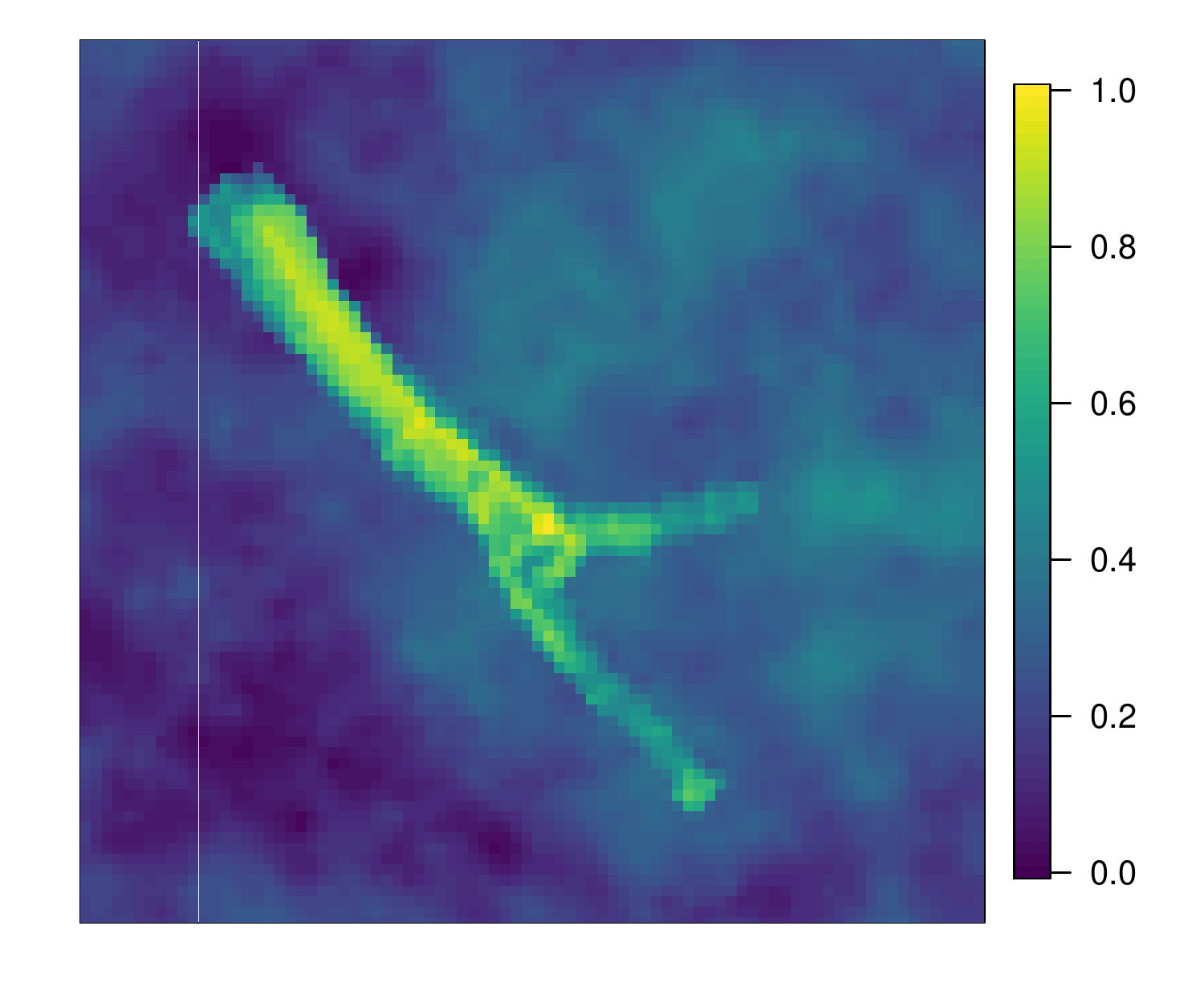}\label{fig:template}}
	\subfloat[Synthetic observation]{\includegraphics[width=0.3\textwidth]{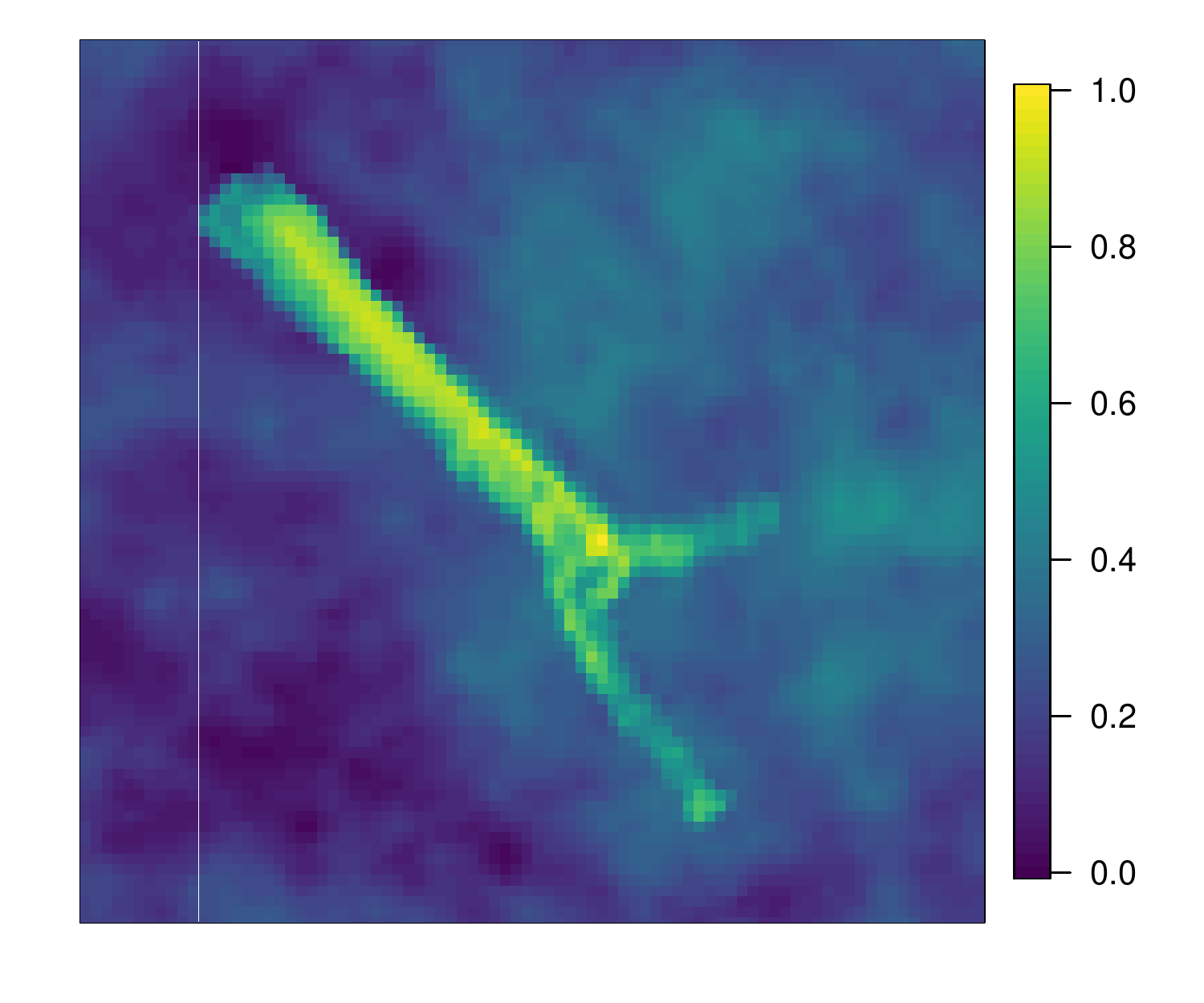}\label{fig:Y_line_all_ex2_obs}}
	
	\subfloat[0.28,  0.45]{\includegraphics[width=0.25\textwidth, page=1]{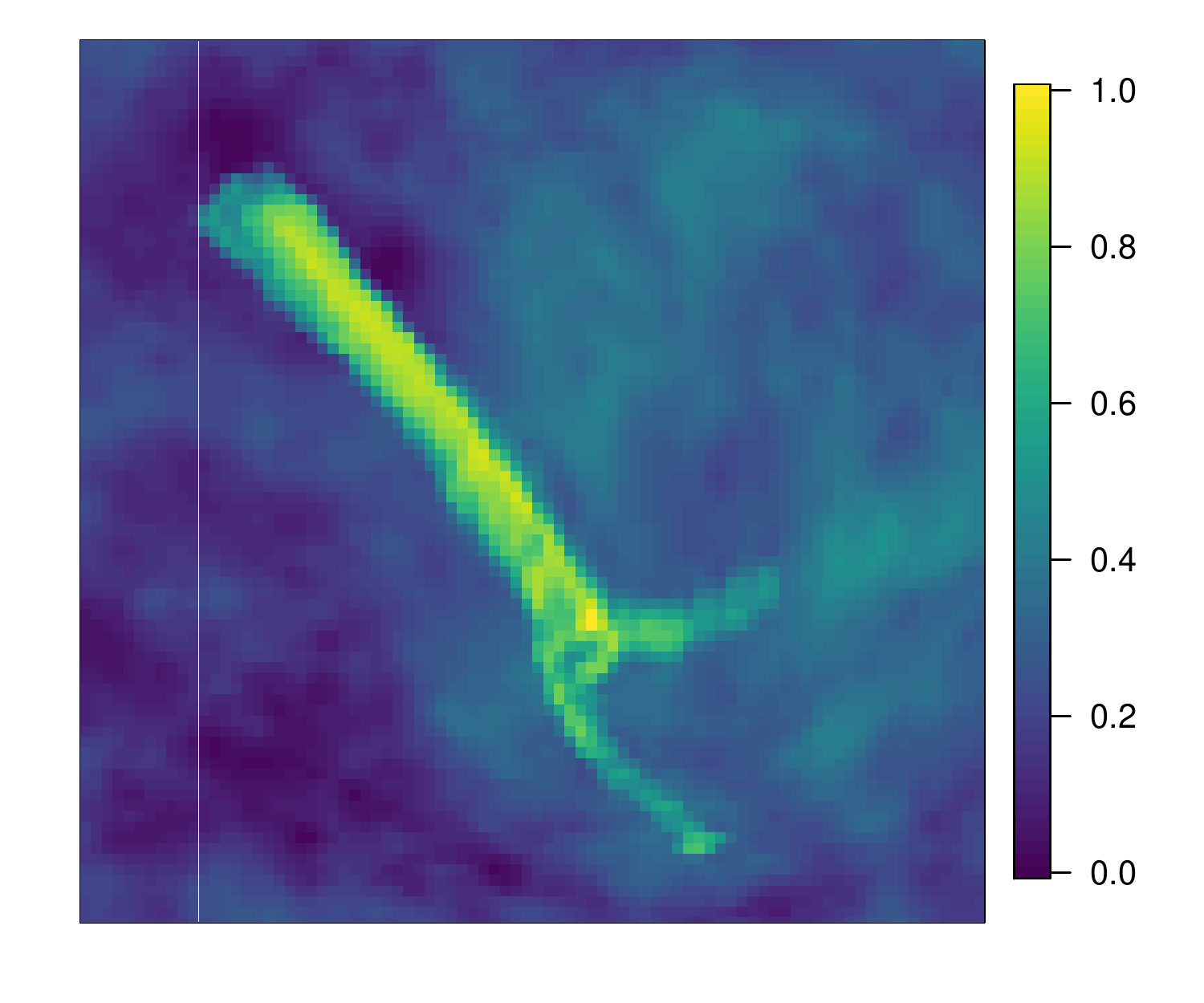}}
	\subfloat[0.36, -0.42]{\includegraphics[width=0.25\textwidth, page=15]{{fig_line_ex2}}}
	\subfloat[-0.10, 0.31]{\includegraphics[width=0.25\textwidth, page=38]{{fig_line_ex2}}}
	\subfloat[-0.37 -0.27]{\includegraphics[width=0.25\textwidth, page=16]{{fig_line_ex2}}}

	\subfloat[warped (c)]{\includegraphics[width=0.25\textwidth, page=1]{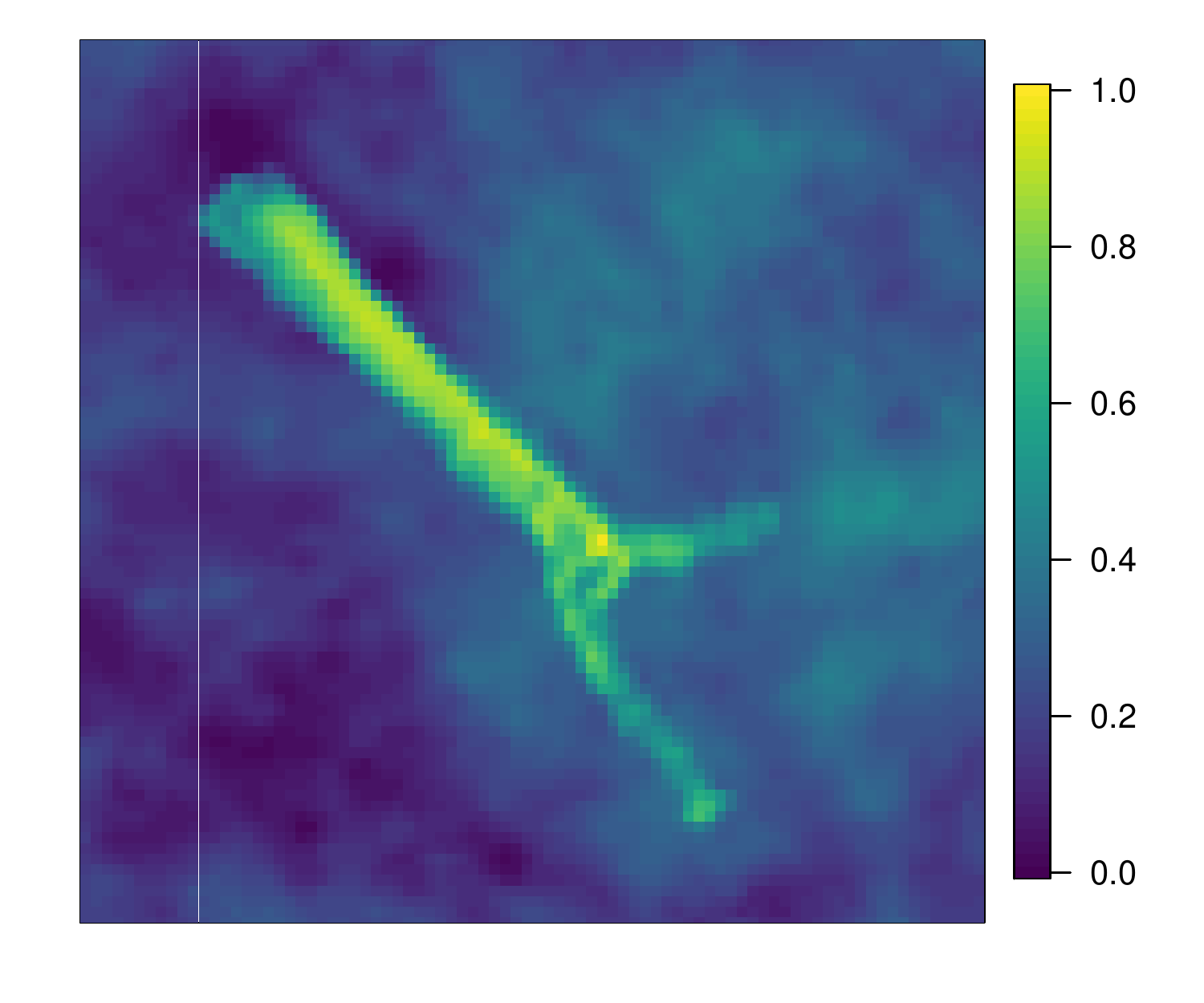}}
	\subfloat[warped (d)]{\includegraphics[width=0.25\textwidth, page=15]{{figWarped_ex2}}}
	\subfloat[warped (e)]{\includegraphics[width=0.25\textwidth, page=38]{{figWarped_ex2}}}
	\subfloat[warped (f)]{\includegraphics[width=0.25\textwidth, page=16]{{figWarped_ex2}}}

    \subfloat[$\gamma$ (c)]{\includegraphics[width=0.25\textwidth,trim={40pt 0 35pt 0},clip]{{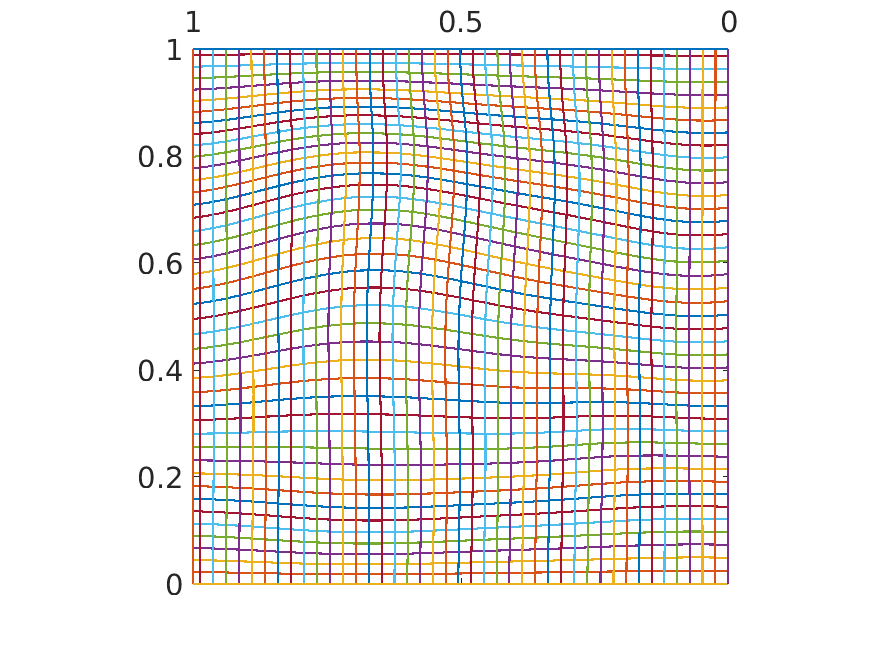}}}
	\subfloat[$\gamma$ (d)]{\includegraphics[width=0.25\textwidth,trim={40pt 0 35pt 0},clip]{{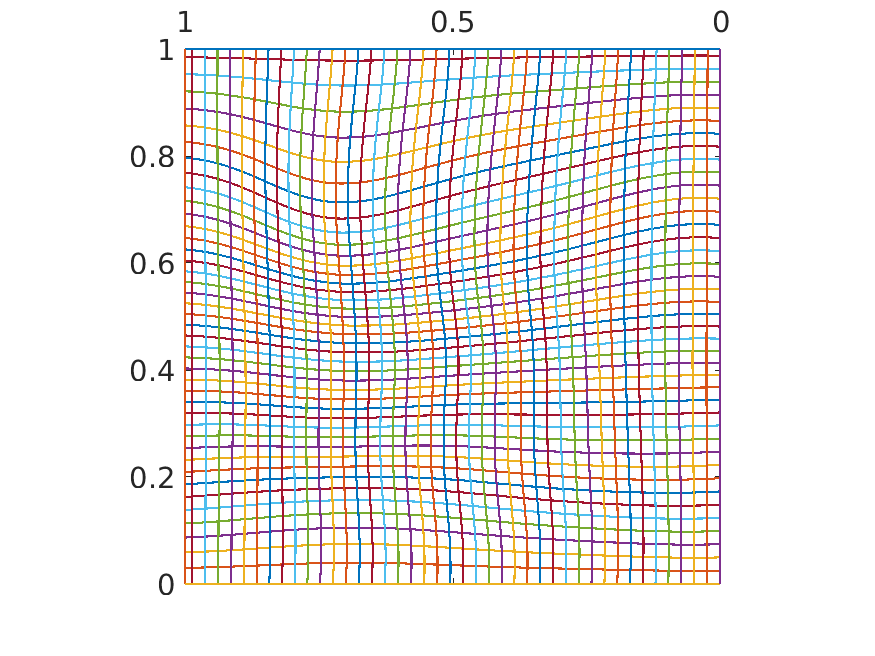}}}
	\subfloat[$\gamma$ (e)]{\includegraphics[width=0.25\textwidth,trim={40pt 0 35pt 0},clip]{{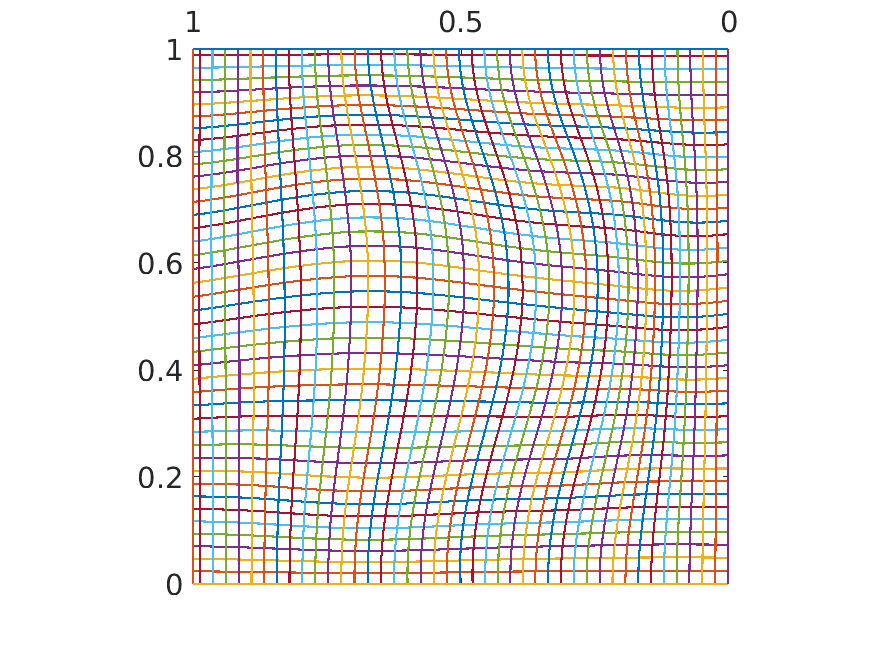}}}
	\subfloat[$\gamma$ (f)]{\includegraphics[width=0.25\textwidth,trim={40pt 0 35pt 0},clip]{{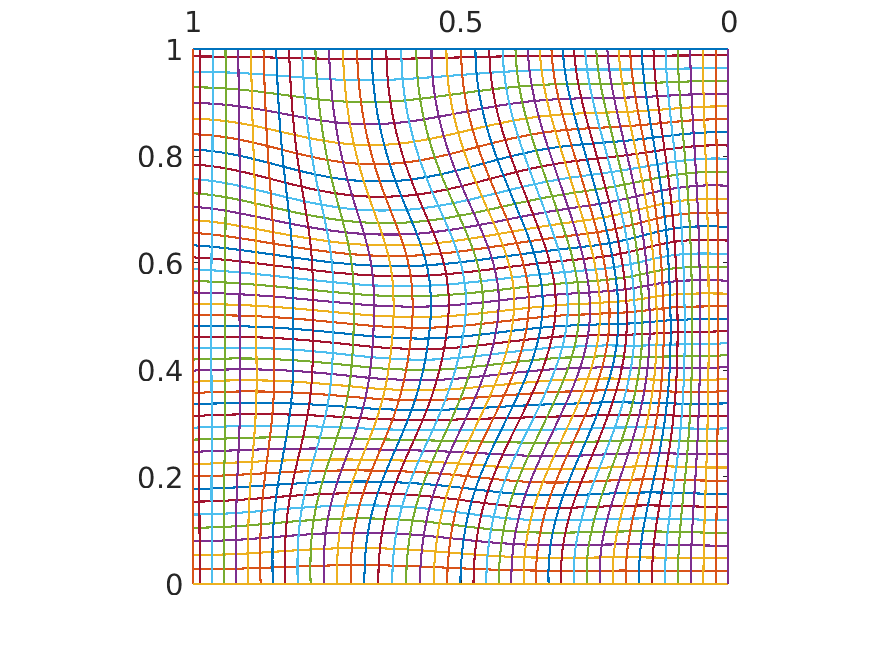}}}

	\caption{Template image (a) for generating model runs and synthetic observation (b). Four examples of model output are shown in Figure (c)-(f), their corresponding warped images (g)-(j) and grid deformations are shown (k)-(n).}\label{fig:Y_line_all_ex2}
	\centering
\end{figure}

We generated 50 model runs at a set of design points chosen using Latin hypercube sampling. Figure \ref{fig:template} shows the 
template image for generating the model runs. For each design point, the template is deformed to create a warped image based on equation \eqref{eqn:warp}. As an illustration, Figure \ref{fig:Y_line_all_ex2} shows the model outputs for four different input parameter settings. The branch located at the bottom right corner of the template image is stretched to the right when $\theta_1$ is positive and compressed to the left when negative, while a similar deformation action is applied to the vertical direction controlled by $\theta_2$. We also generate a synthetic observation (Figure \ref{fig:Y_line_all_ex2_obs}) from the diffeomorphism warp using input parameter $(0.3,0.1)$. Therefore, all the model runs here have misaligned and misshapen features compared to the synthetic observation.\\

We compute the deformation metrics for each model run using the synthetic observation as the reference. {\color{black} In this case our image is of the form $F(s_1,s_2)=\nabla f=(\partial_{s_1}f,\partial_{s_2}f)$ where $f(s_1,s_2)$ is the pixel intensity. We calculate the gradient using finite difference approximations.} The optimal deformations for the four example runs and the warped model runs are shown in Figure \ref{fig:Y_line_all_ex2}. We see that all four warped model runs are indistinguishable to the synthetic observation visually. This indicates the deformation has been applied correctly and estimated well.\\

\begin{figure}[hpb]
	\centering
	\subfloat[amplitude and phase]{\includegraphics[width=0.4\textwidth,page=7,trim={0 0 0 12pt},clip]{fig_line_post}}
	\subfloat[amplitude]{\includegraphics[width=0.4\textwidth,page=8,trim={0 0 0 12pt},clip]{fig_line_post}}

	\subfloat[phase]{\includegraphics[width=0.4\textwidth,page=9,trim={0 0 0 12pt},clip]{fig_line_post}}
	\subfloat[Euclidean]{\includegraphics[width=0.4\textwidth,trim={0 0 0 12pt},clip]{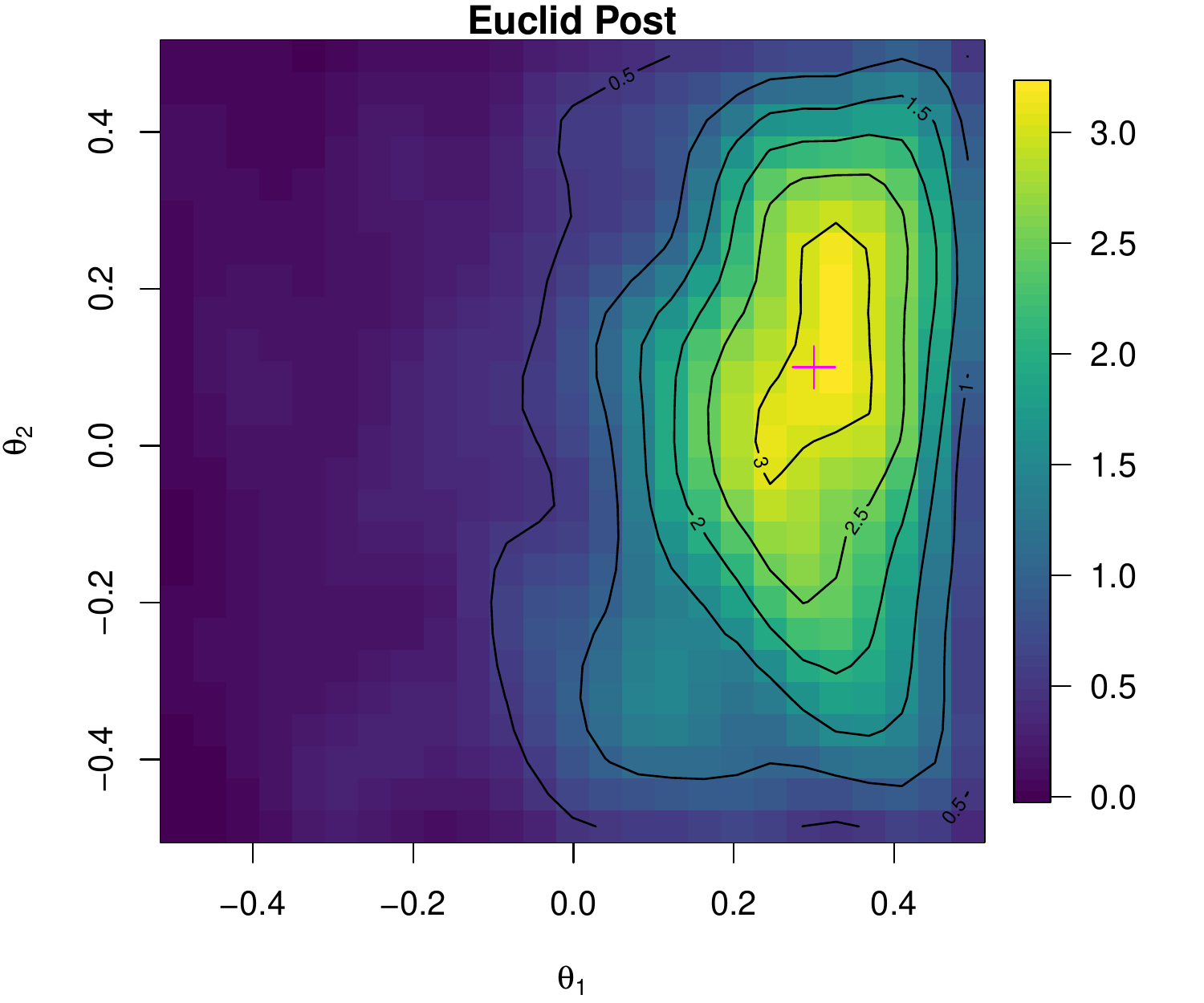}}
	\caption{\textbf{Single crack toy example:} Posterior density function of input parameters using both amplitude and phase, amplitude only, phase only and Euclidean distance. Magenta color ``+" indicates the input parameters corresponding to the synthetic observation.}
	\label{fig:fig_line_kde_ex2}
\end{figure}

We fit independent GP emulators to each metric then perform Bayesian calibration as described in Section \ref{s:cali}. {\color{black} In our MCMC implementation, we sample three 30,000 posterior samples at different initial values with 15,000 burn-in; MCMC convergence is assessed using the Gelman-Rubin diagnostic \citep{coda}}. Calibration results using different metrics are presented in Figure \ref{fig:fig_line_kde_ex2}. It is notable that the posterior distributions from calibration using the amplitude or Euclidean alone are very flat, providing little information on the input parameter, while the posterior using the phase distance peaks near the true input parameter value, and the calibration using the combined metrics is similar to using phase distance alone. The posterior modes and credible intervals are: for amplitude and phase $\theta_1= 0.36 (0.08 ,0.49), \theta_2=0.10 (-0.08,0.28)$ , for amplitude only $\theta_1= 0.19 (-0.27,0.49), \theta_2=-0.04 (-0.34,0.49)$, for phase only $\theta_1= 0.39 (0.10,0.49), \theta_2=0.10 (-0.13, 0.33)$, and for euclidean $\theta_1= 0.34(-0.19,0.49), \theta_2=0.08(-0.45, 0.45)$.
\\

The results confirm that the proposed method is able to properly account for misaligned and misshapen model outputs. Because the model runs and the synthetic observation for this toy example are all warped from the same template, all model runs have the same amplitude distance except for small numerical differences due to interpolation in generating the images. This means there exists a $\gamma^*$ that exactly maps each image to the synthetic observation, therefore, the phase distance quantifies the amount of warping in order to correct for misaligned and misshapen features. The calibration using the phase distance alone suggests the input parameter setting that produces the correct geometric features which are least misaligned and misshapen. This is particularly useful for model development and evaluation as a measure of a model's ability to produce correct coherent structures, a geometric feature that persists in a flow field. 
Moreover, for models which produce highly localized coherent structures, hurricanes for instance, the Euclidean metric may completely disqualify a model run which produced the correct hurricane characteristics just in the wrong location. In this case the model itself may be well calibrated with location error coming from external sources, such as error in an initial condition taken from observations. 

\subsection{Sea Ice Model Experiment}\label{s:experiment_ice}
{\color{black} In the previous toy example our model runs were generated by an actual diffeomorphism, however in most geophysical applications model output comes from a physical numerical model. These model outputs corresponding to different parameters may not be an exact diffeomorphism warp of each other.} Here we perform a perfect model experiment for the sea ice model described in Section \ref{s:data}. {\color{black} In this perfect model experiment we take the observation to be the model output for a chosen set of parameters.}  {\color{black} This experiment demonstrates, through the recovery of the chosen model parameterization using the calibration scheme,} that the method works well for physical models whose output is not generated from a diffeomorphism. 

 We produce 80 models runs for a set of Latin hypercube sampled parameters by varying the tensile and shear strength parameters in the input space $\bTheta=[10,50]\times[10,150]$. Each model run provides gridded output which we map to the unit square, our domain $D$.  The output is the jump vector $[\mathbf{u}](\bs), \bs=(s_1,s_2) \in D$  which tracks the orientation and size of a crack as discussed in Section \ref{s:data}. From this we obtain the opening magnitude (magnitude of the vector) and the orientation ($\alpha$) of the crack. Model outputs are shown in the Appendix. 
Using these values we can define our image to be a function $f: \mathbb{R}^2 \rightarrow \mathbb{R}^n$ where $n=2,3$ corresponding to images defined as $f(\bs)=(\left|[\mathbf{u}]\right|(\bs),\alpha(\bs))$, $f(\bs)=(\partial_{s_1} \left|[\mathbf{u}]\right|(\bs),\partial_{s_2} \left|[\mathbf{u}]\right|(\bs))$, or $f(\bs)=(\partial_{s_1} \left|[\mathbf{u}]\right|(\bs),\partial_{s_2} \left|[\mathbf{u}]\right|(\bs), \alpha(\bs))$. {\color{black} Here the gradients of the magnitude of the jump vectors are taken using a finite difference approximation.} The choice of which image type to analyze is an important one and should be based on the model considered. For our perfect model experiment we find that the true parameters are discovered by calibration, posterior mode location, for each image type, however differences in posterior shapes are observed. This should be expected as the images themselves are in fact different. Considering only the magnitude and not orientation, for example, calibration based on the combined metrics produces a multimodal posterior suggesting a larger set of ``acceptable" model parameters. We have found that several model runs produce similar opening magnitudes in similar locations. This could mean that ice crack magnitudes and locations are not overly sensitive to the sampled parameters in some regions of the parameter space. Therefore, the orientation of the ice cracks provides additional information that is useful for calibration.\\

For our perfect model experiment we choose to focus on the image defined by $f(\bs)=(\partial_{s_1} \left|[\mathbf{u}]\right|(\bs),\partial_{s_2} \left|[\mathbf{u}]\right|(\bs), \alpha(\bs))$. We find that the gradient of magnitude provides a better alignment of features than considering magnitude alone due to the extra information as input.  Calibration results for image defined by $f(\bs)=( |[\mathbf{u}]|(\bs), \alpha(\bs))$ are presented in the Appendix for comparison. \\

To carry out the analysis, we compute the optimal warping function for each model run. Assessing the metrics corresponding to the model runs, we notice that the phase distances are left skewed. As a result, we compare cross-validation performance on GP emulators fitted to the original and the log transformed distances; results indicate that data transformation is necessary for this experiment to ensure the surrogate model has good prediction performance, for instance, small mean square prediction error and reasonable coverage. \\

\begin{figure}[hpt]
	\centering
	\subfloat[amplitude and phase]{\includegraphics[width=0.4\textwidth,page=3,trim={0 0 0 12pt},clip]{fig_ice_post}}
	\subfloat[amplitude]{\includegraphics[width=0.4\textwidth,page=4,trim={0 0 0 12pt},clip]{fig_ice_post}}

	\subfloat[phase]{\includegraphics[width=0.4\textwidth,page=5,trim={0 0 0 12pt},clip]{fig_ice_post}}
	\subfloat[Euclidean]{\includegraphics[width=0.4\textwidth,trim={0 0 0 12pt},clip]{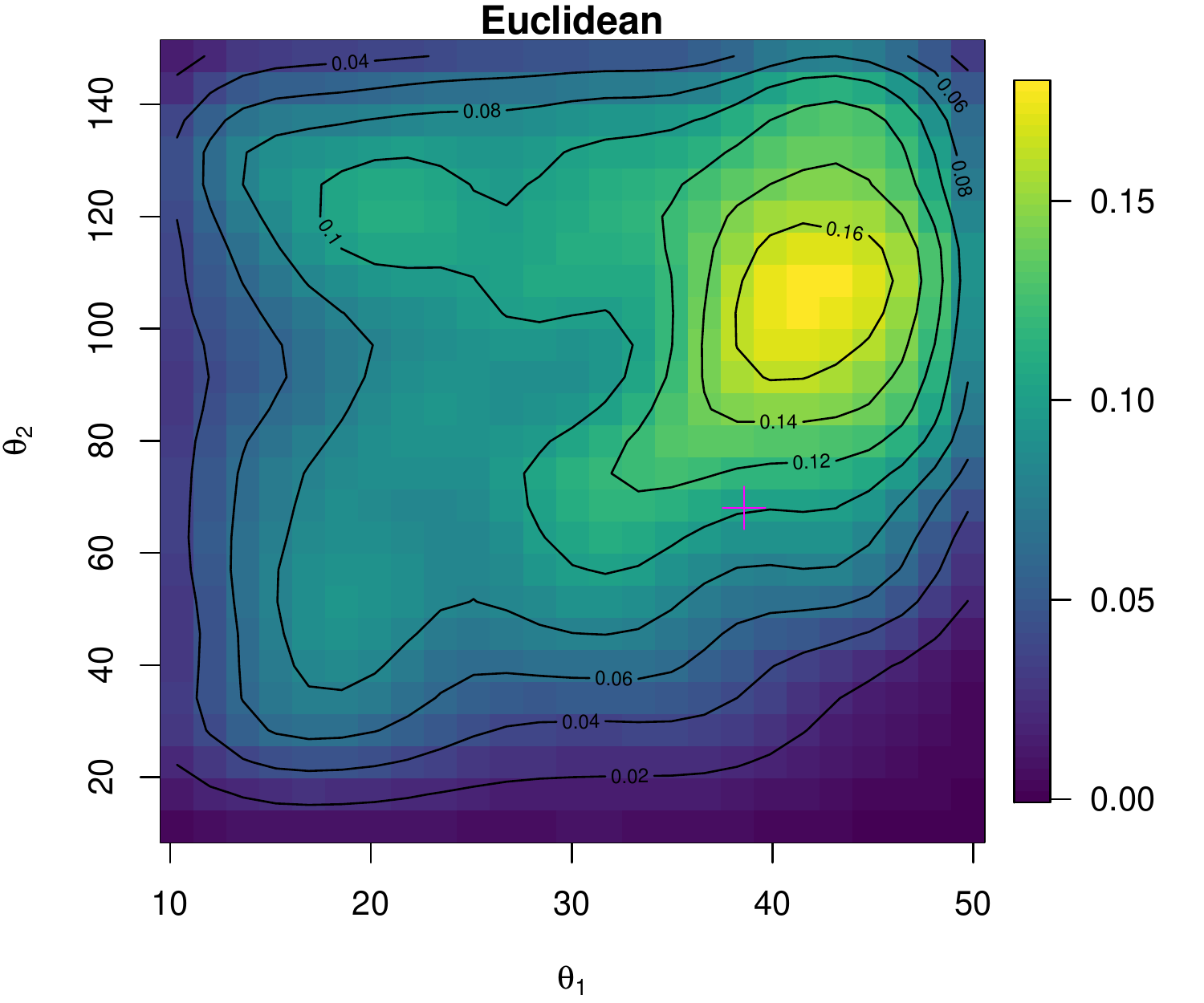}}

	\caption{\textbf{Perfect model experiment:} Posterior density function of input parameters using both amplitude and phase, amplitude only, phase only and Euclidean distance. Magenta color ``+" indicates the input parameters corresponding to the synthetic observation.}
	\label{fig:fig_ice}
\end{figure}

{\color{black} Bayesian calibration and the necessary convergence diagnostic are then carried out;} the posterior density functions are shown in Figure \ref{fig:fig_ice}. The marginal posterior mode and credible interval for different metrics are summarized in Table \ref{tab:inference_experiment}. Calibration using both amplitude and phase metrics has posterior mode closest to the true value with the smallest credible interval length compared to other metrics. Calibration using Euclidean or amplitude alone provides relative flat posterior with a wrong mode. However, we have not added additional observation errors to the synthetic observation in this perfect experiment. In the presence of complicated error in observations, the calibration method for inference should incorporate the sophisticated model error mechanism.

\begin{table}[hpt]
    \centering
    \caption{\textbf{Inference:} Marginal posterior modes and credible intervals using different metrics.}
    \label{tab:inference_experiment}.
    \begin{tabular}{|c|cc|}
    \hline
    & tensile strength (38.581) & shear strength (68.032)\\
    \hline
    Amplitude and phase &45.714 ( 32.904 , 49.697 ) & 77.193 ( 68.633 , 94.472 )\\
    Amplitude &42.513 ( 21.358 , 49.696 ) & 83.529 ( 68.479 , 99.638 )\\
    Phase &45.577 ( 20.886 , 49.686 ) & 77.197 ( 66.026 , 95.769 )\\
    Euclidean &41.608 ( 13.306 , 49.631 ) & 89.624 ( 66.68 , 99.392 )\\
    \hline
    \end{tabular}
\end{table}

\section{Application to Sea Ice Model and Sattelite Data}\label{s:appli}
\begin{figure}[hpt]
	\centering
	\subfloat[amplitude and phase]{\includegraphics[width=0.4\textwidth,page=3,trim={0 0 0 12pt},clip]{fig_ice_post_rgps}}
	\subfloat[amplitude]{\includegraphics[width=0.4\textwidth,page=4,trim={0 0 0 12pt},clip]{fig_ice_post_rgps}}

	\subfloat[phase]{\includegraphics[width=0.4\textwidth,page=5,trim={0 0 0 12pt},clip]{fig_ice_post_rgps}}
	\subfloat[Euclidean]{\includegraphics[width=0.4\textwidth,trim={0 0 0 12pt},clip]{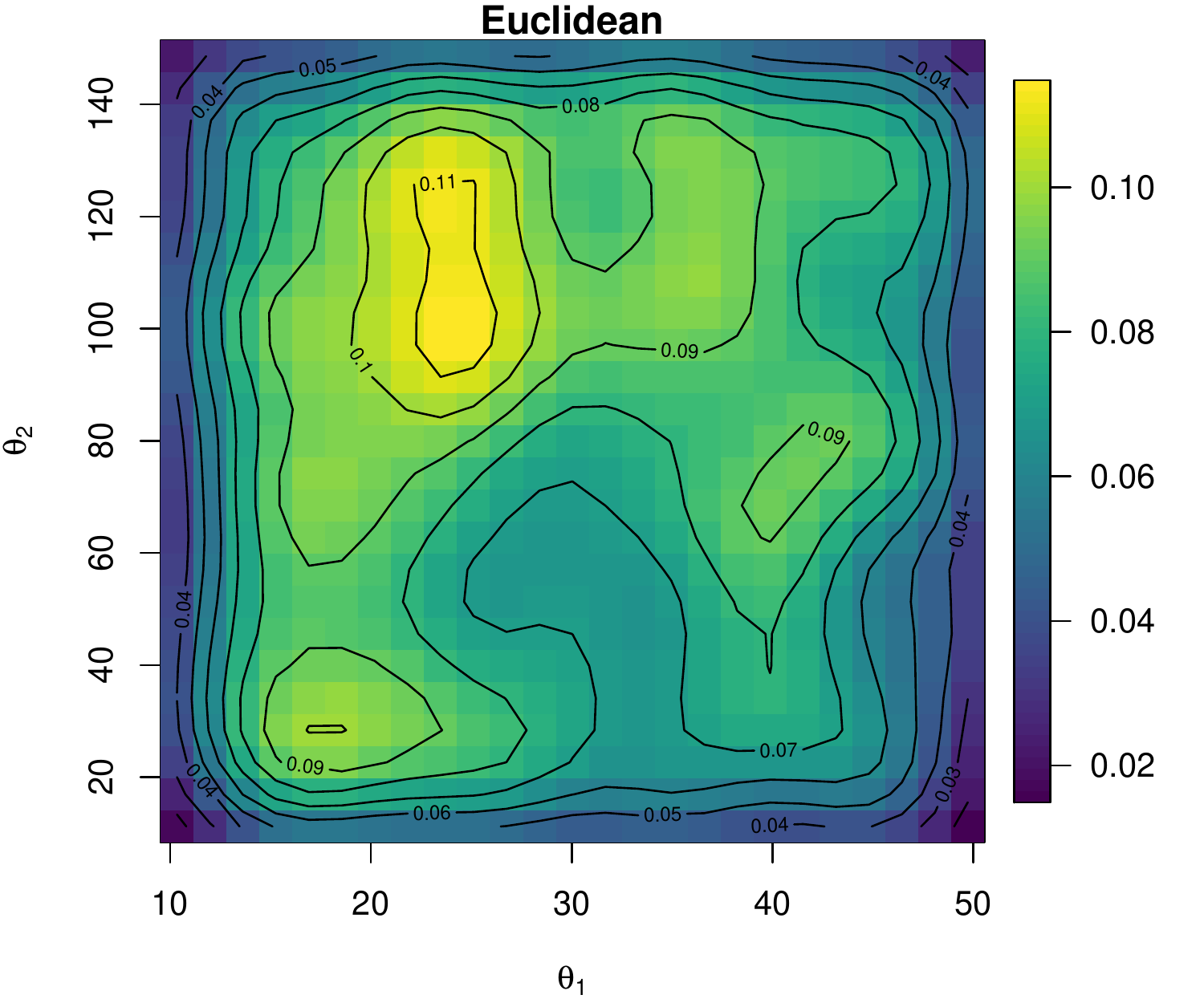}}

	\caption{\textbf{RGPS data application:} Posterior density function of input parameters using both amplitude and phase, amplitude only, phase only and Euclidean distance.}
	\label{fig:fig_ice_rgps}
\end{figure}

\begin{figure}[ht]
    \centering
    \subfloat[]{\includegraphics[width=0.35\textwidth]{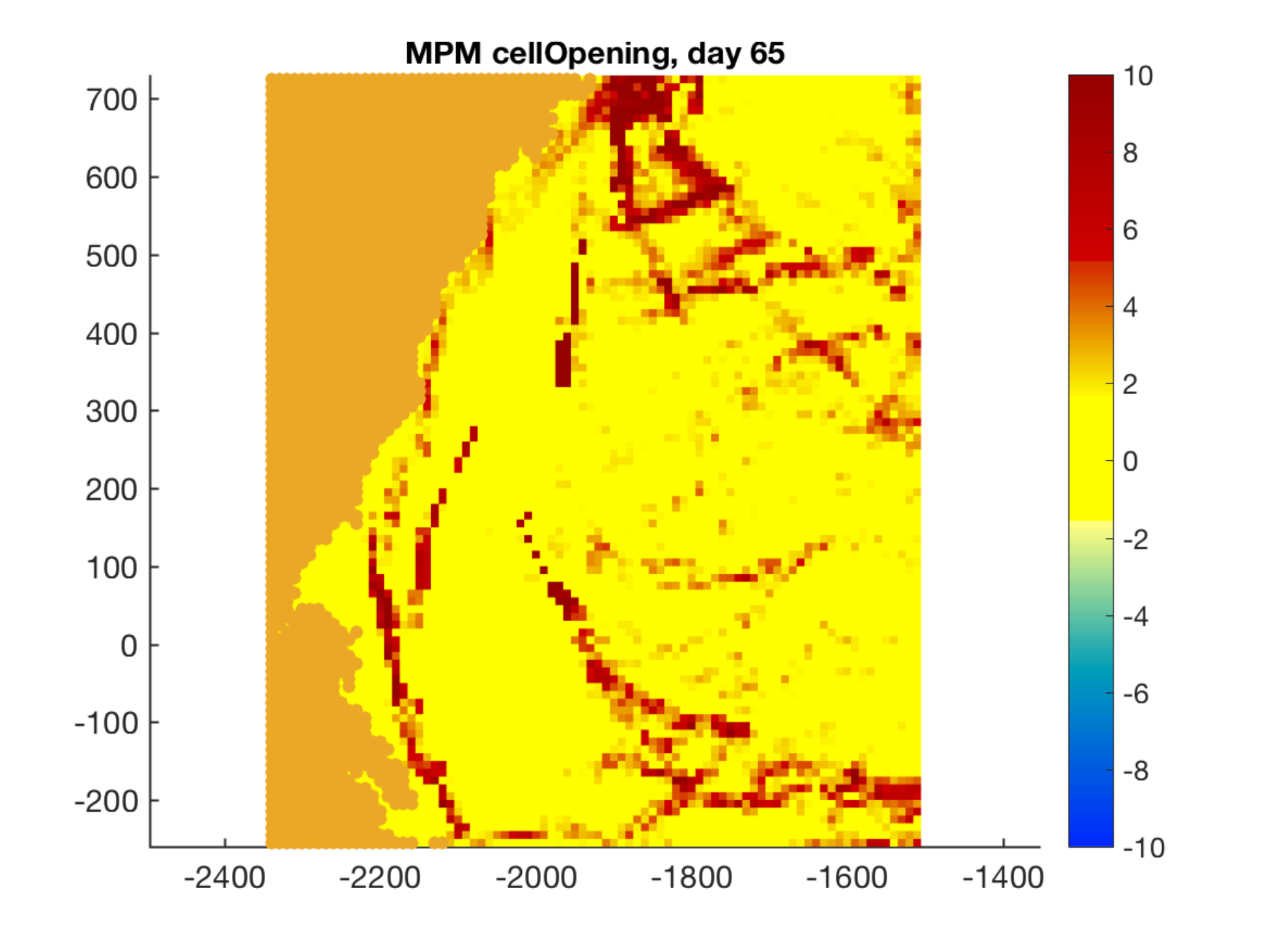}}
     \subfloat[]{\includegraphics[width=0.35\textwidth]{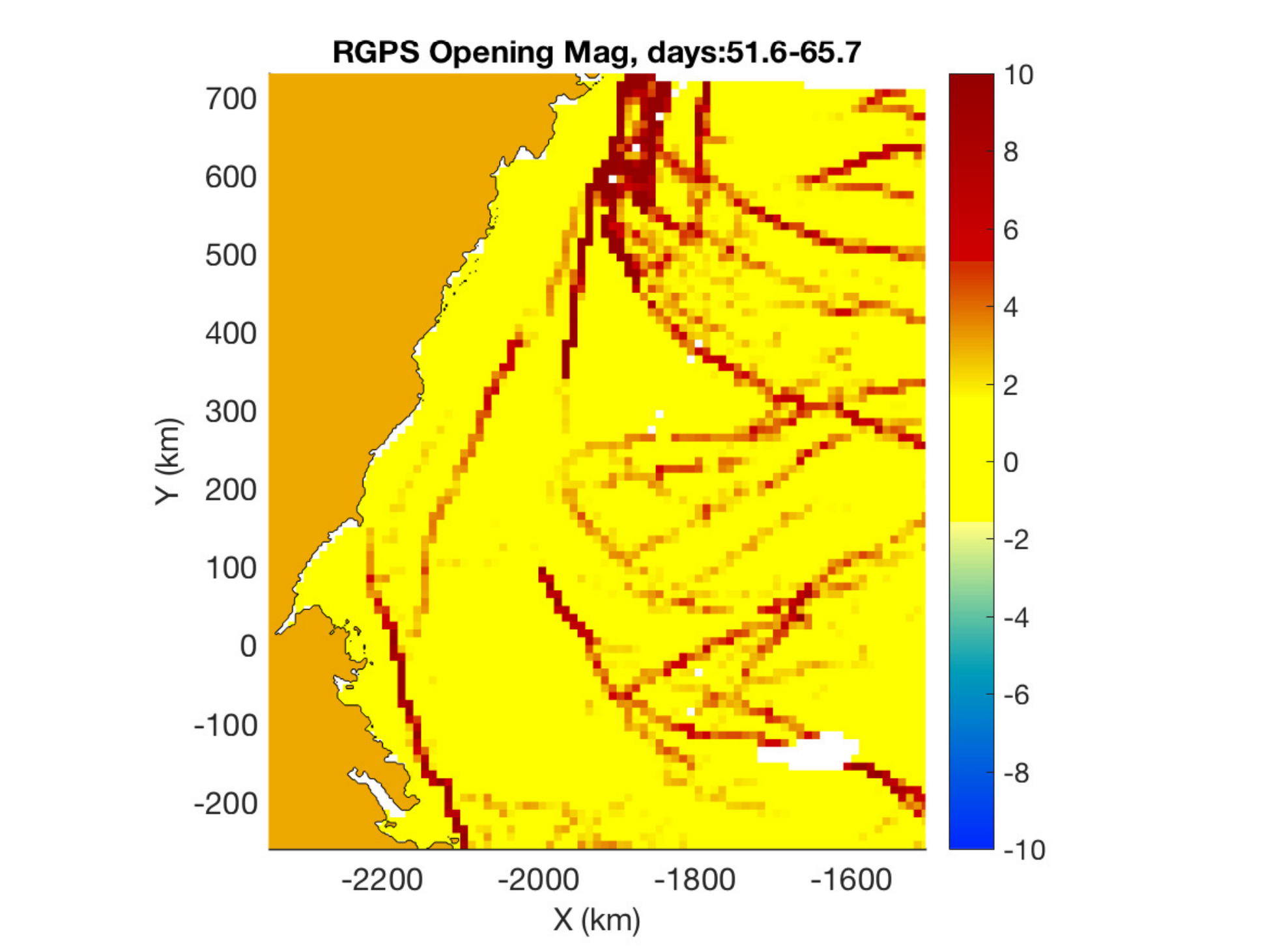}}
    \subfloat[]{\includegraphics[width=0.35\textwidth]{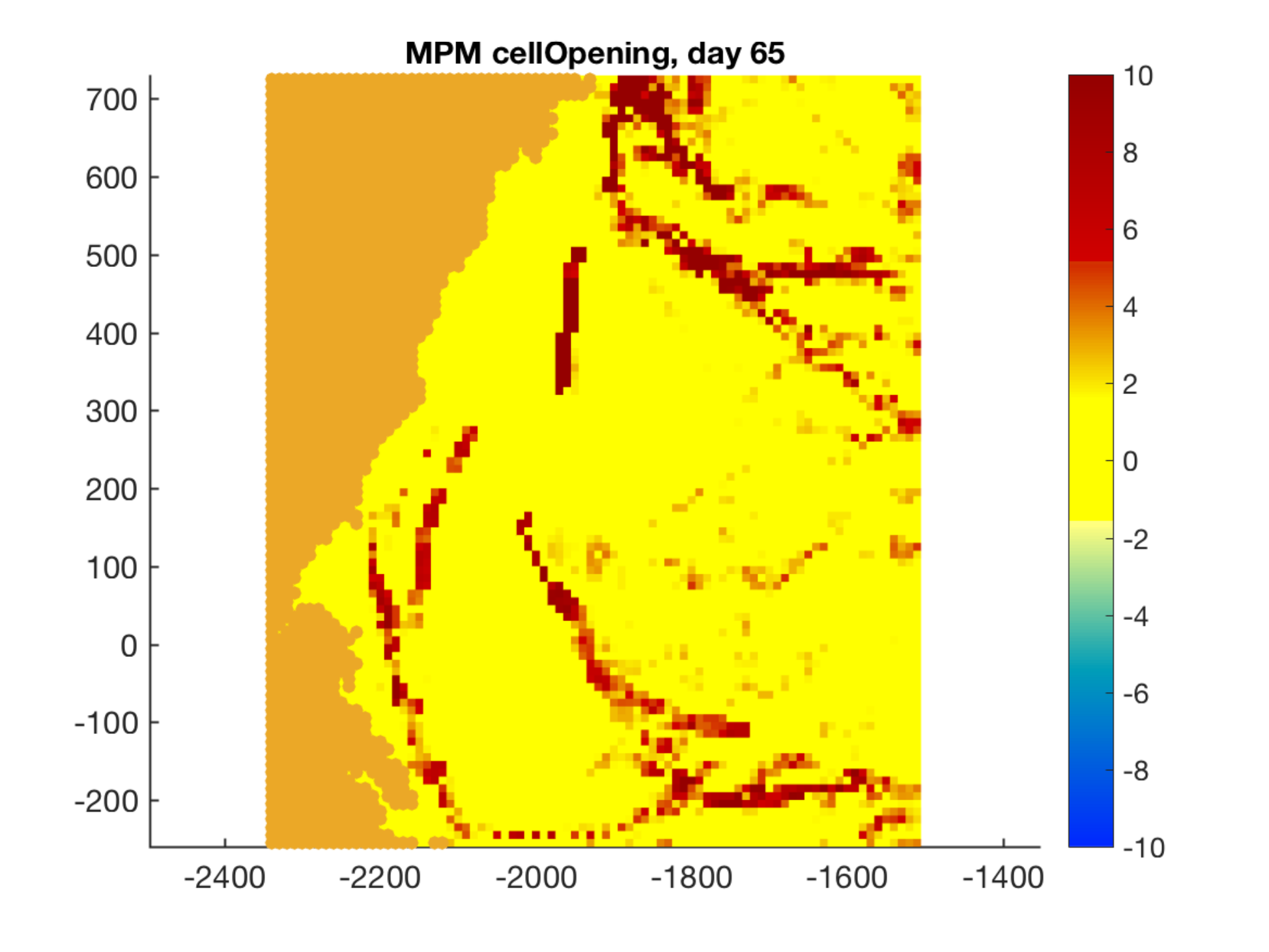}} \\
     \subfloat[]{\includegraphics[width=0.35\textwidth]{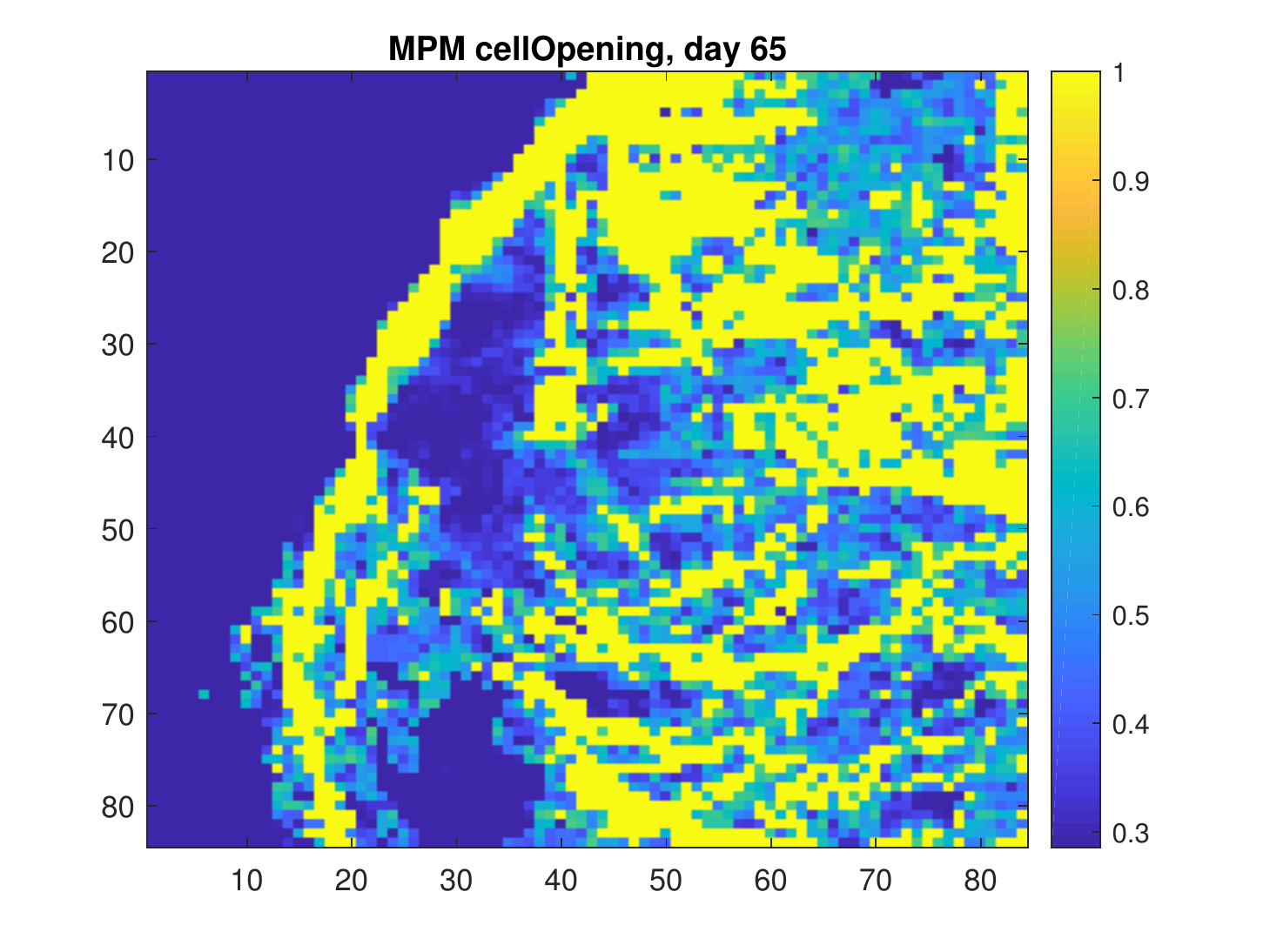}}
     \subfloat[]{\includegraphics[width=0.35\textwidth]{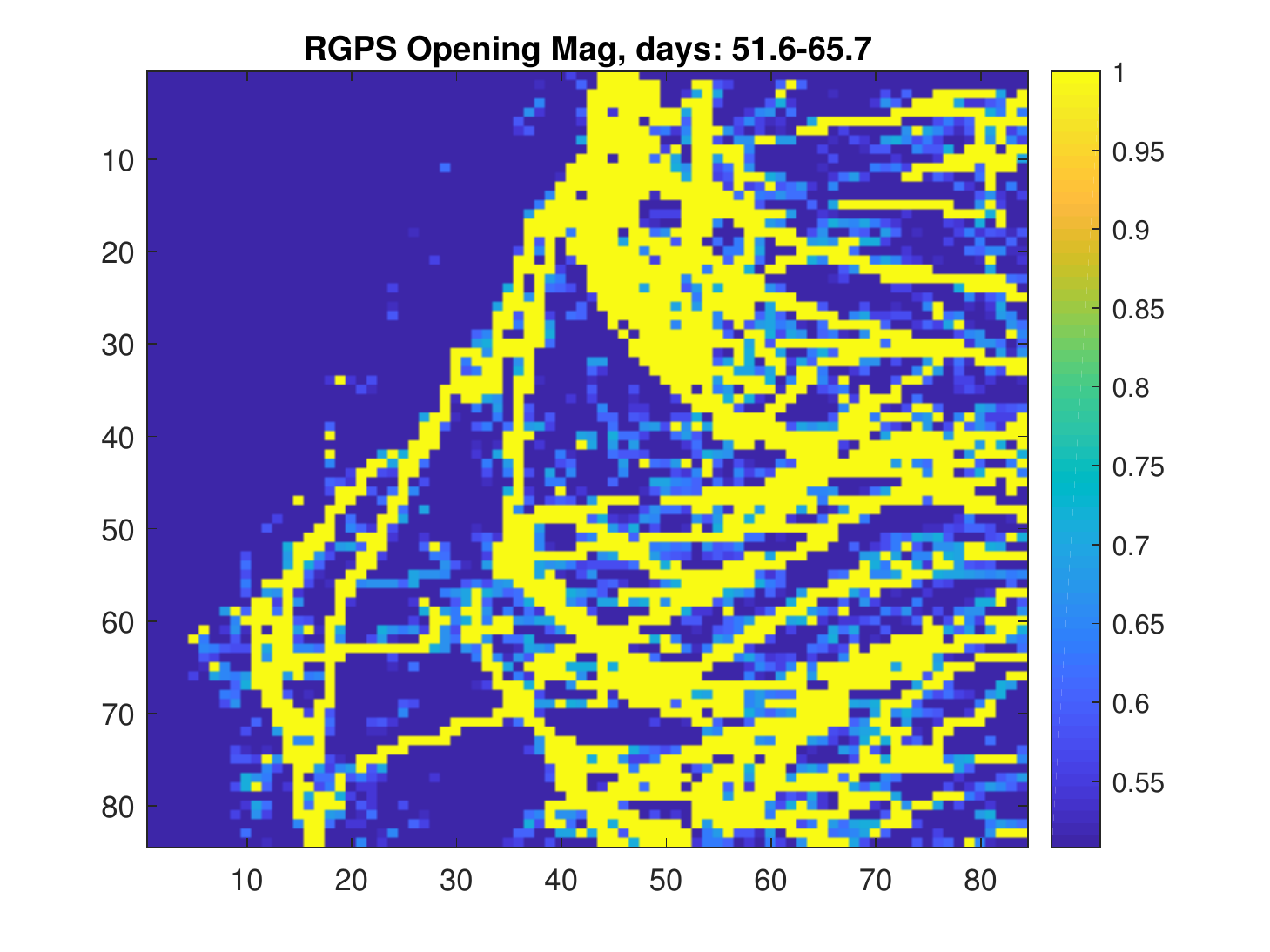}}
    \subfloat[]{\includegraphics[width=0.35\textwidth]{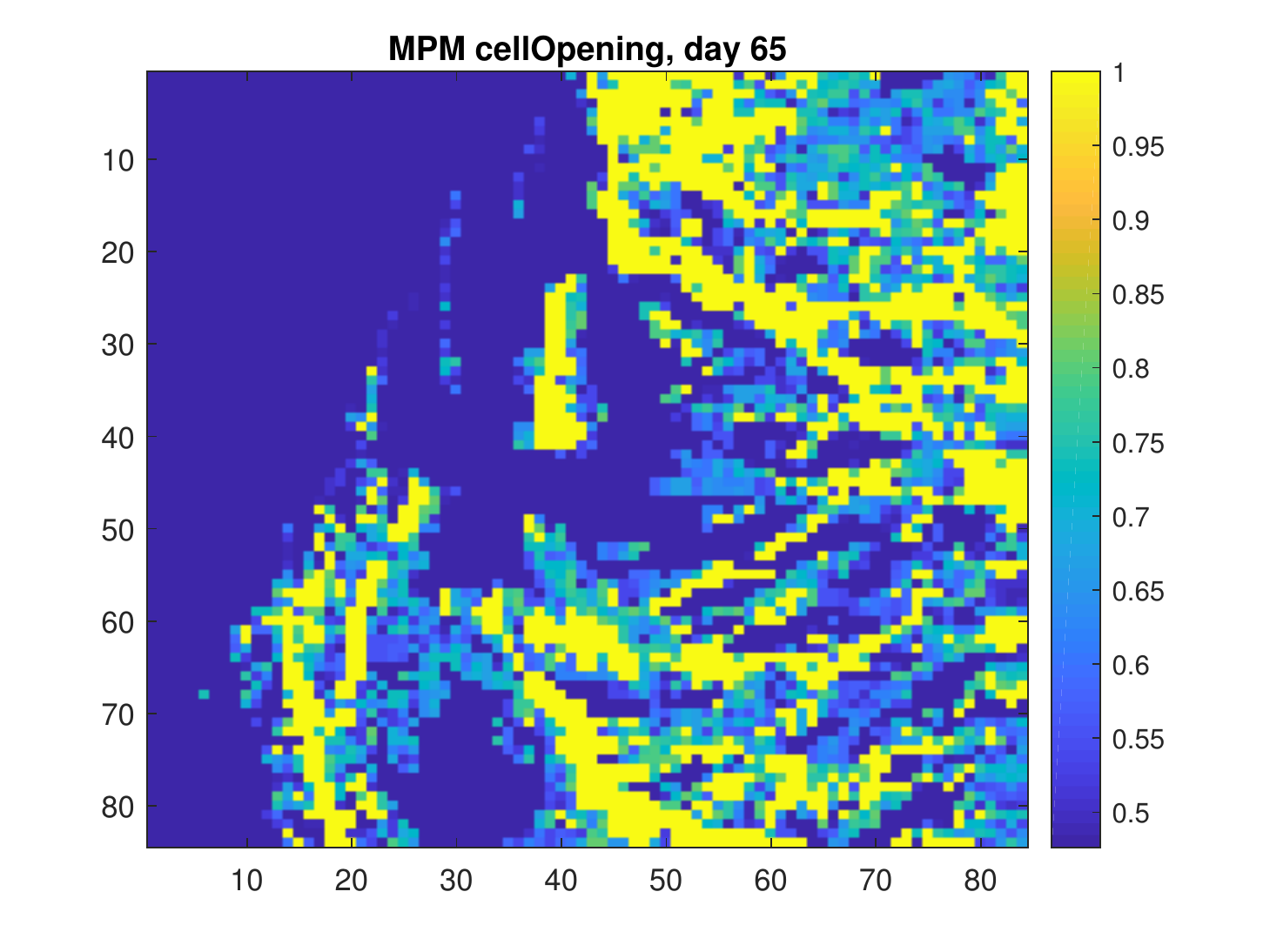}}
    \caption{Comparison of magnitude of jump in displacement (a) simulation without calibration using parameters in \cite{SulskyPeterson2011}, (b) RGPS opening magnitude from the kinematic algorithm, and (c) simulation using the calibrated parameters. In (d)-(e) we show the same data as (a)-(c) but with contrast enhanced to highlight differences.}
    \label{fig:RGPSCal}
\end{figure}

The proposed method is applied to the RGPS data discussed in Section \ref{s:data}. We warp each model output to RGPS data using the same choice for step size and the number of basis functions to represent $\gamma$ as in the sea ice experiment in Section \ref{s:experiment_ice}. {\color{black} MCMC convergence has been assessed using the Gelman - Rubin diagnostic \citep{coda} on three MCMC chains with different initial values, where each chain is obtained from drawing 30,000 posterior samples and discarding the first 15,000 for burn-in.} Calibration results are presented in Figure \ref{fig:fig_ice_rgps}. {\color{black} The posterior modes for the tensile and shear strength parameters from using the combined metrics are 25.45 (11.601, 48.063) and 75.71 (60.514, 95.42), respectively, with the credible intervals given in parenthesis}. To establish that the posterior mode represents a viable parameterization, we run the elastic-decohesive model and assess the output with input parameters equal to the posterior mode, shown in Figure \ref{fig:RGPSCal}. {\color{black} In contrast to the perfect model case above, we cannot know whether there is a set of parameters which can exactly reproduce the RGPS data, in fact likely not. To quantify the performance of the model with the calibrated parameters we compare both the Euclidean and q-map distances (i.e. $\mathbb{L}^2$ distance between q-maps of the images) between the model runs and the RGPS data, for both the calibrated parameter settings and the original parameter settings chosen from physical considerations in \cite{SulskyPeterson2011}. We also apply our warping algorithm using each model output to measure the amplitude and phase distances between the runs and the RGPS data. Before preforming any warping we find that the parameter settings used in \cite{SulskyPeterson2011} give q-map and Euclidean distances of 23.9 and 46.3, respectively, between the model run and observations compared to that of 18.2 and 42.4 for the model run using the calibrated parameters. The amplitude distances for both model runs are similar at 11.2 for the original parameters and 11.9 for the calibrated parameters. The most striking difference comes from the phase distances at 5.4 for the original parameters and 1.1 for the calibrated parameters. This suggests much less warping was required to align geometric features when comparing the calibrated model run to the observations. }

We believe the calibration based on both amplitude and phase provide a more reliable inference on input parameters than using Euclidean alone based on the perfect model study where the truth is known. In the instance shown in Figure \ref{fig:fig_ice} the Euclidean distance is less informative with a very diffuse posterior and a mode which misses the true value.  Further, the center mode in Figure \ref{fig:fig_ice_rgps} encompasses that chosen from physical considerations in \cite{SulskyPeterson2011}. In the Appendix we show the case where our image was taken to be $f(\bs)=(|[\bu]|(\bs),\alpha(\bs))$.

\section{Discussion}\label{s:discussion}
The calibration method presented is the first of its kind based on the image warping metrics. This method is particularly useful for model calibrations where capturing coherent structures is of importance. Further, the metric itself is a proper distance provided the optimal $\gamma$ is found. Currently, we are using gradient descent, local minima are a possibility. The gradient decent scheme can be very sensitive to step size and the number of basis functions chosen to represent $\gamma$. In this work, we manually tuned the step size and number of basis functions to produce an optimal result. Future work would include looking at other optimization methods. The proposed method also reduces the high-dimensional model outputs into scalar responses, such that a one-dimensional Gaussian process model can easily handle the emulation problem. Warping algorithms are implemented in MATLAB, while emulation and calibration are implemented in R \citep{R}.\\

We illustrated the usefulness of the proposed method through a toy example and a perfect model experiment using the elastic-decohesive model. {\color{black} We used 50 design points for the first experiment and 80 for the latter. While the number of design points required depends on the complexity of the response surface and the number of active inputs influencing the response, the sample size we used here was sufficient for our experiments. A rule of thumb for selecting the number of design points is 10d, where d is the dimension of the input space \citep{Loeppky2009,Harari2018}}. While the results in our experiments show that the proposed method performs well, we note that we have not added observation error when generating the synthetic observation. This can be a major caveat where a more rigorous investigation should be conducted. It remain future work to study how observation error could affect warping and calibration results.\\

The ultimate application of this method should be to real world observations. {\color{black} The use of image warping techniques is motivated by a need to assess sea ice model output consisting of fractures. For climate modeling, it is not necessary to get these features exactly right.  In fact, it is not likely that  we  would  be  able  to  predict the  cracks  exactly since  the  forcing  that drives  the  motion  in  the model  is  too  uncertain.   Thus, we  desire  a  metric  that  can  compare  data  and  model  favorably  if  the predicted cracks can be warped to the observed cracks that is, if the predicted cracks are somewhat misaligned or misshapen.} In our study using the satellite-derived RGPS data of sea ice deformation we are able to obtain informative posteriors using the combination of amplitude and phase metrics, and the posterior mode seems to represents a viable parameterization, when compared to that chosen from physical considerations in \cite{SulskyPeterson2011}.  

\section*{Acknowledgements}
This material was based upon work partially supported by the National Science Foundation under Grant DMS-1638521 to the Statistical and Applied Mathematical Sciences Institute and by the National Oceanic and Atmospheric Administration under grant NA15OAR4310165 to the University of New Mexico. MH was also partially supported by NSF-DMS1418090 and through the Network for Sustainable Climate Risk Management (SCRiM) under NSF cooperative agreement GEO-1240507. Any opinions, findings, and conclusions or recommendations expressed in this material are those of the authors and do not necessarily reflect the views of the National Science Foundation, U.S. Department of Energy, or the United States Government.


\bibliographystyle{plainnat}
\bibliography{refs}

\section*{Appendix A.1: Model Outputs}
\begin{figure}[hpt]
	\centering
	\includegraphics[width=\textwidth]{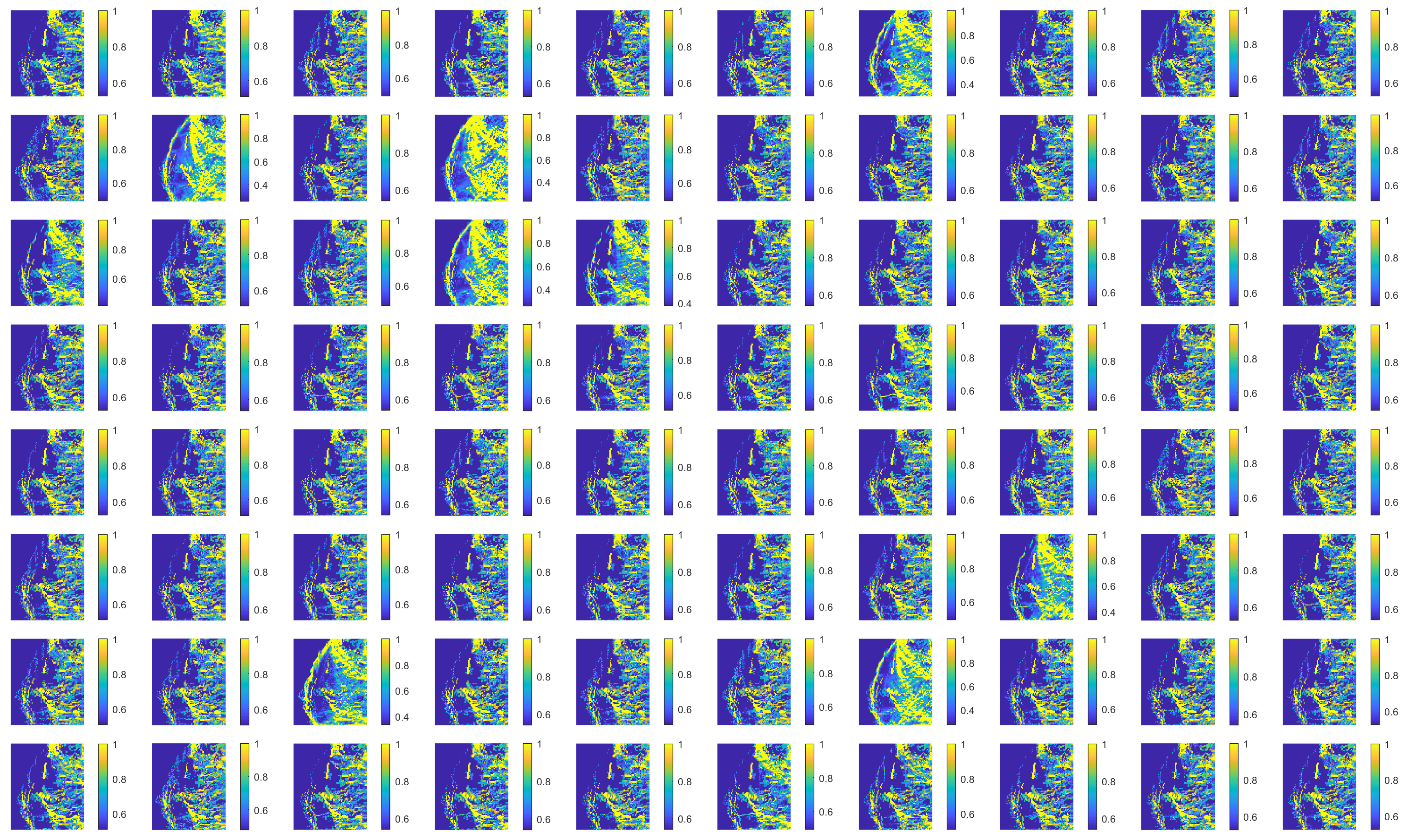}
	\caption{Magnitude of the jump vectors corresponding to the 80 model runs}
	\label{fig:jump}
\end{figure}

\begin{figure}[hpt]
	\centering
	\includegraphics[width=\textwidth]{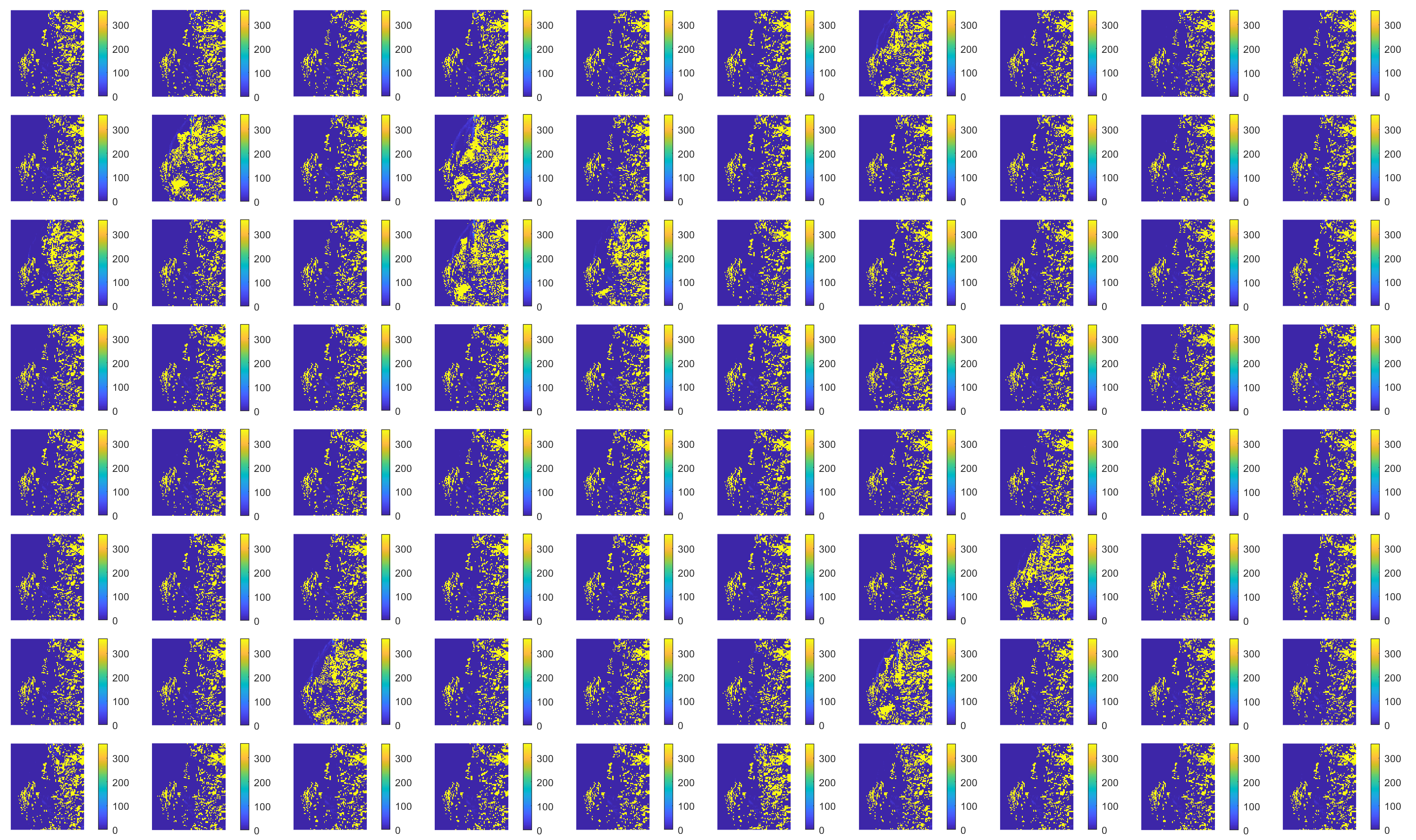}
	\caption{Angles of the jump vector corresponding to the 80 model runs}
	\label{fig:angle}
\end{figure}

We have obtained 80 model runs by varying the tensile and shear strength parameters. The model outputs here are contrast enhanced to help with visualizing the ice features. We see that the several model runs have similar magnitude (Figure \ref{fig:jump}). This could be an indication that magnitude at current simulation time step and simulation period is not sensitive to some regions of input parameter space. Orientations of ice cracks from the model runs are shown in Figure \ref{fig:angle}. Incorporating these model output in calibration method is useful for parameter inference.


\section*{Appendix A.2: Image Selection}
\begin{figure}[hpt]
	\centering
	\subfloat[amplitude and phase]{\includegraphics[width=0.34\textwidth,page=3,trim={0 0 0 12pt},clip]{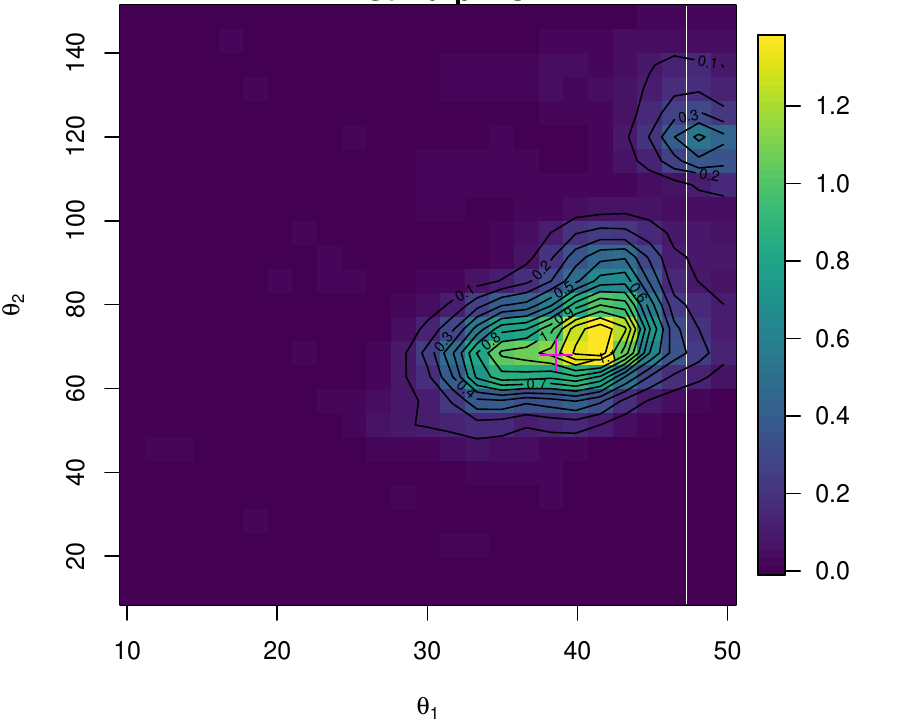}}
	\subfloat[amplitude]{\includegraphics[width=0.34\textwidth,page=4,trim={0 0 0 12pt},clip]{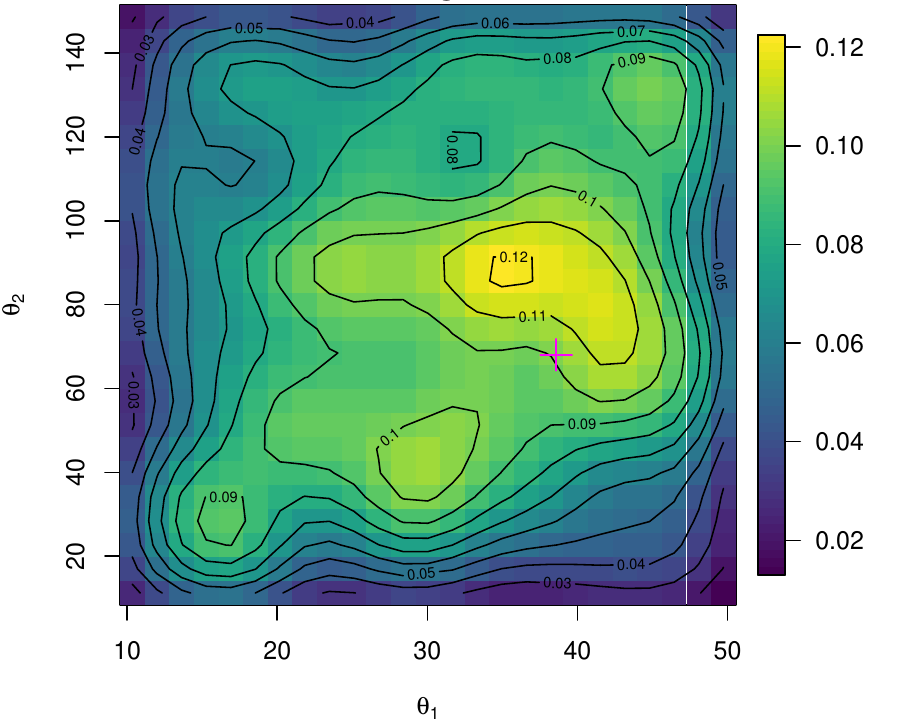}}
    \subfloat[phase]{\includegraphics[width=0.34\textwidth,page=5,trim={0 0 0 12pt},clip]{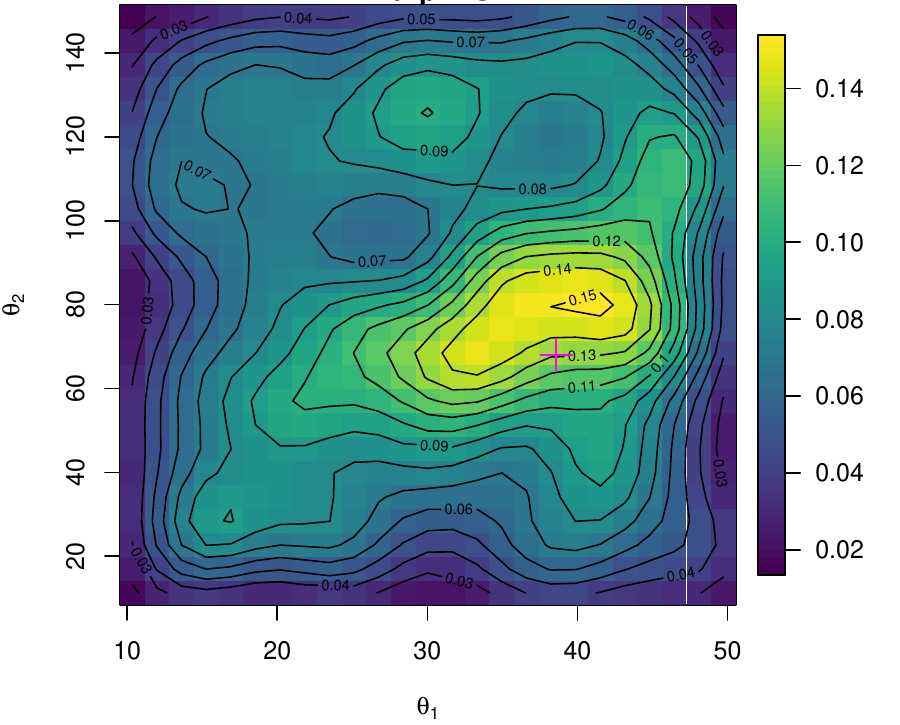}}\\

	\centering
	\subfloat[amplitude and phase]{\includegraphics[width=0.34\textwidth,page=3,trim={0 0 0 12pt},clip]{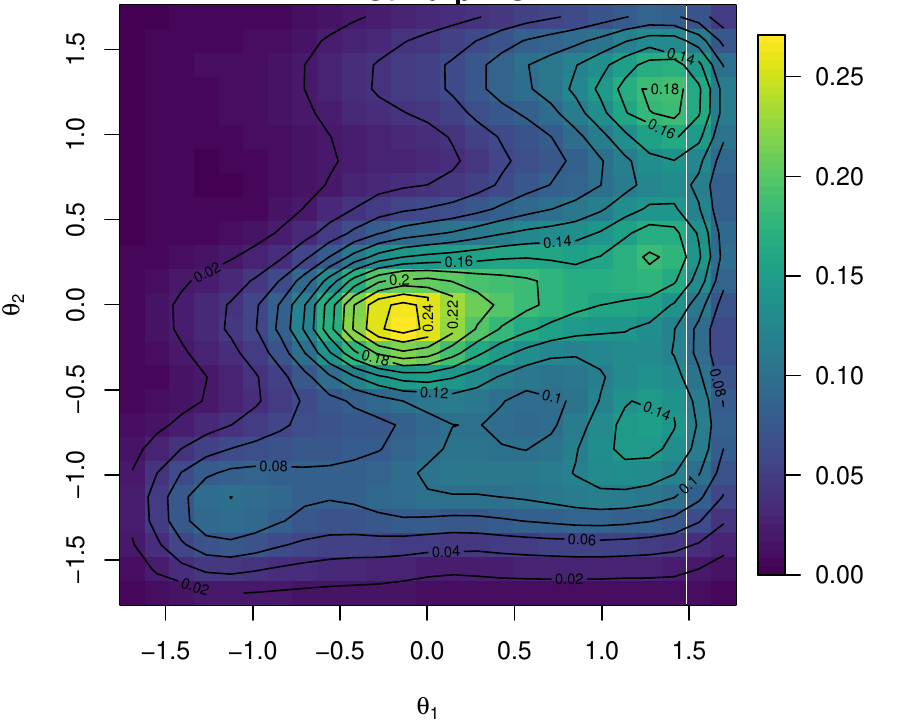}}
	\subfloat[amplitude]{\includegraphics[width=0.34\textwidth,page=4,trim={0 0 0 12pt},clip]{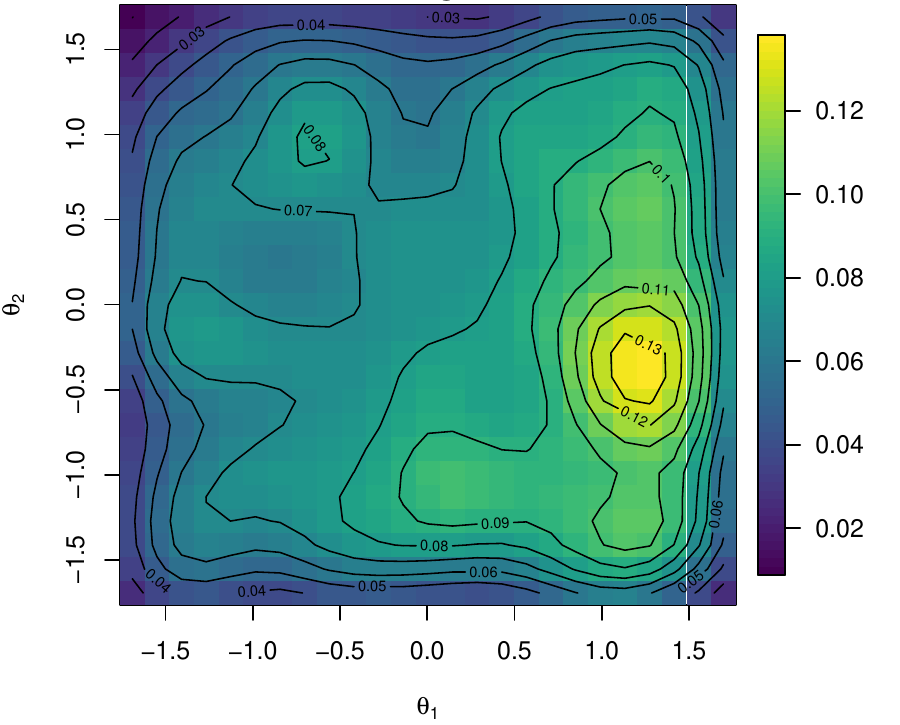}}
	\subfloat[phase]{\includegraphics[width=0.34\textwidth,page=5,trim={0 0 0 12pt},clip]{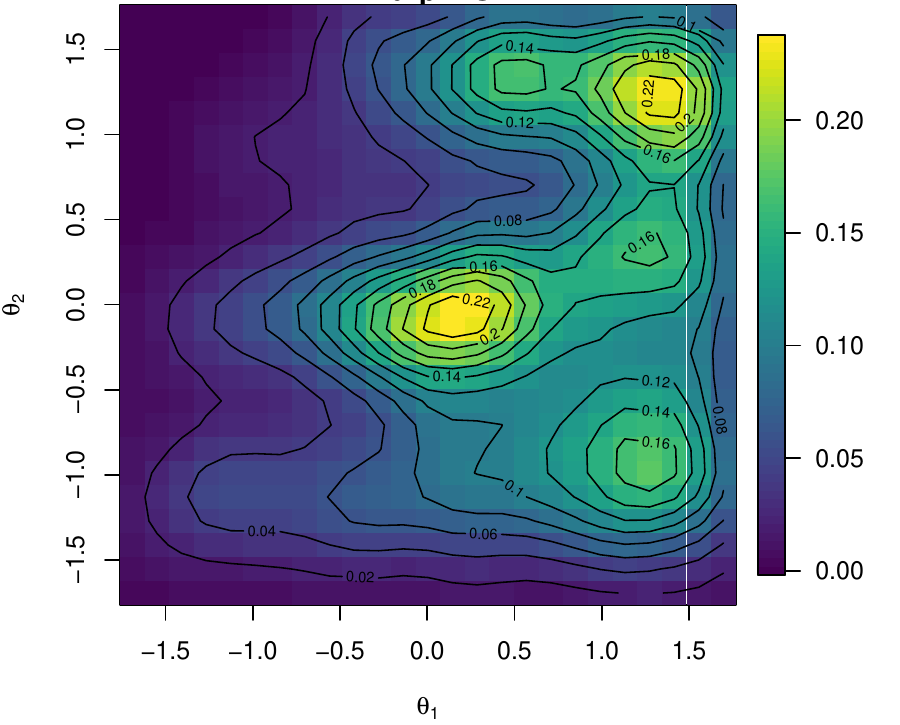}}
	\caption{Calibrations using no gradients. Top: \textbf{Perfect model experiment:} Posterior density function of input parameters using both amplitude and phase, amplitude only and phase only. Magenta color ``+" indicates the input parameters corresponding to the synthetic observation. Bottom: \textbf{RGPS data application:} Posterior density function of input parameters using both amplitude and phase, amplitude only, and phase only.}
	\label{fig:fig_ice_mag_angle_rgps}
\end{figure}

In Figure \ref{fig:fig_ice_mag_angle_rgps}, we present the calibration results for image defined as $f(s_1,s_2)=(|[u]|,\alpha)$ for the sea ice experiment and data application.  We found through the perfect model study that the image type $f(s_1,s_2)=(\partial_{s_1}|[u]|,\partial_{s_2}|[u]|,\alpha)$ typically provided better results. This extra information provided by the gradient gave our method with the ability to better distinguish between model predictions. It is notable that for our perfect model case the Euclidean distance failed to pick up the the true value at the posterior mode while the warping metrics did. The calibration using no gradient for the RGPS data application showed multiple modes that were ruled out when using the gradient of the opening magnitudes.  

\end{document}